\documentclass{aa}  

\usepackage{graphicx}
\usepackage{lipsum}
\usepackage{txfonts}

\usepackage{float}
\usepackage{newfloat}
\usepackage{hyperref}
\hypersetup{colorlinks, linkcolor={blue}, citecolor={blue}, urlcolor={blue}}
\usepackage{afterpage}
\usepackage{amsmath}
\usepackage{subcaption}
\usepackage{siunitx}
\usepackage{booktabs}
\usepackage{multirow}
\usepackage{color}
\usepackage{xcolor} 
\usepackage{tcolorbox}

\definecolor{raspberry}{rgb}{0.7,0.,0.3}
\definecolor{magenta}{rgb}{0.8,0,0.8}
\definecolor{purple}{rgb}{0.5,0,0.5}
\definecolor{gray}{rgb}{0.5,0.6,0.7}
\usepackage{silence}
\usepackage{enumitem}

\begin{document}

   \title{eROSITA cosmology with galaxy groups: Hot gas budget out to the virial radius}

   \author{H. Khalil\thanks{Corresponding author: hossam.khalil@helsinki.fi} \inst{1}
          \and
          A. Finoguenov \inst{1}
        \and
        D. Eckert\inst{2}
        \and
        R. Seppi\inst{2}
        \and
        E. Tempel\inst{3,4}
        \and
        L. Lovisari\inst{5,6}
        \and
        F. Gastaldello\inst{7}
        }

   \institute{Department of Physics,  University of Helsinki, Gustaf Hällströmin katu 2A, Helsinki, FI-00014, Finland 
    \and
    Department of Astronomy, University of Geneva, Ch. d’Ecogia 16, CH-1290 Versoix, Switzerland      
    \and 
    Tartu Observatory, University of Tartu, Observatooriumi 1, 61602 Tõravere, Estonia
    \and
     Estonian Academy of Sciences, Kohtu 6, 10130 Tallinn, Estonia
    \and
      INAF-IASF Milano, Via Alfonso Corti 12, 20133 Milan, Italy
    \and
       Center for Astrophysics | Harvard \& Smithsonian, 60 Garden Street, Cambridge, MA 02138, USA
    \and INAF, Istituto di Astrofisica Spaziale e Fisica Cosmica di Milano, via A. Corti 12, 20133 Milano, Italy   
             }

   \date{Received \today}

\abstract {Non-gravitational processes that expel hot gas beyond the virial regions of groups and clusters of galaxies, known collectively as baryonic feedback, play a key role in reshaping the matter distribution of the Universe on megaparsec scales. Tracing the baryonic content of galaxy groups outside $R_{500}$ provides a unique opportunity to constrain the modification of the matter distribution on these scales. Our aim is to quantify the baryon budget at the virial radii of galaxy groups and constrain the effect of active galactic nuclei feedback in redistributing matter on the largest scales ever probed with X-ray observations. We used eROSITA observations of a complete sample of 25 galaxy groups selected from the first public release of the eROSITA-DE data (eRASS1) and identified with the Two Micron Redshift Survey optical group catalogue (2MRS). We extracted and fitted surface brightness (SBx) profiles and present hot gas mass and hot gas fraction profiles out to $R_{200}$. We explored the physical properties of our sample by performing a Bayesian analysis of $M_{\mathrm{gas}}-M_{\mathrm{tot}}$, $L_{\mathrm{X}}-M_{\mathrm{tot}}$, and $L_{\mathrm{X}}-M_{\mathrm{gas}}$ relations, taking into account the aperture covariance effects. At $R_{500}$, we report uniformly flat SBx profiles with a mean $\beta$ parameter of $0.38 \pm0.04$, steepening to $\beta = 0.76\pm0.19$ beyond $R_{500}$. We measure a sub-cosmic hot gas fraction at the median mass of our sample, $M_{500} = 2.54\times10^{13}M_{\odot}$, of $ f_{\mathrm{ gas,500}} = 4.32\pm0.42\%$. Similarly, at $R_{200}$ and the median mass, $M_{\mathrm{ 200}} = 3.69\times10^{13}M_{\odot}$, we obtain $f_{\mathrm{gas,200}}=5.78\pm0.69\%$. Our $f_{\mathrm{gas}}-M_{\mathrm{tot}}$ and $L_{\mathrm{X}}-M_{\mathrm{tot}}$ relations deviate significantly from predictions of the strong feedback variants of the FLAMINGO simulation ($2.5\sigma$ to $8.0\sigma$ tension), whereas the fiducial FLAMINGO and BAHAMAS models provide the closest match to our measurements. Our results are consistent with prior X-ray observations but achieve twice the statistical precision of recent stacking studies. Using our measured baryon fractions and the hydrodynamical simulation-based model \texttt{SP(k)}, we infer a $10\%-15\%$ reduction in the matter power spectrum at $k = 5\ h\ \mathrm{Mpc}^{-1}$ relative to a dark matter-only universe ($0.85 \leq P(k)/P_{\mathrm{DMO}}(k) \leq 0.90$), in agreement with the predictions of fiducial FLAMINGO and BAHAMAS, while revealing a growing tension on smaller scales with the strong feedback variants.}

\keywords{Galaxies: groups: general -- Galaxies: clusters: intracluster medium -- X-rays: galaxies: clusters -- Cosmology: large-scale structure of Universe}

   \maketitle
   

\section{Introduction}

X-ray emission from the hot gaseous atmospheres of galaxy groups and clusters is a sensitive tracer of the non-gravitational processes that redistribute matter within and around dark matter halos. In the absence of such processes, the hot gas fraction within massive halos would approach the cosmic baryon fraction, $f_{\rm b,cosmic} = \Omega_{\rm b}/\Omega_{\rm m} \approx 0.157$ \citep{Planck2016}, and the gas mass would scale self-similarly with halo mass. However, observations over the past two decades have firmly established that this simple picture is broken: the hot gas fraction within $R_{\rm 500c}$\footnote{The radius enclosing 500 times the critical density of the Universe, $\rho_{\rm c} = 3H^{2}(z)/8\pi G$, at a given redshift, $z$.} of galaxy groups ($M_{\rm 500}= 10^{13}-10^{14}M_{\odot}$) falls significantly below the cosmic value, indicating that a substantial fraction of baryons have been displaced from the inner regions (see \citealt{eckert21} for a review).

\begin{figure*}
    \centering
    \begin{subfigure}[b]{0.45\textwidth}
        \hspace{-0.1in}
            \includegraphics[height=0.28\textheight]{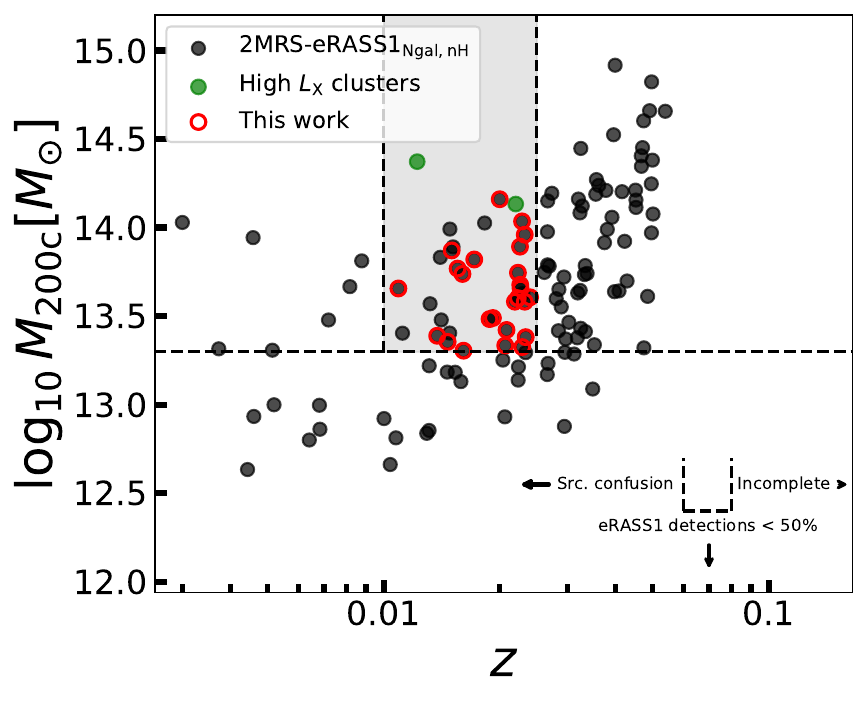}
    \end{subfigure}
    \begin{subfigure}[b]{0.45\textwidth}
            \hspace{0.1in}
            \includegraphics[height=0.28\textheight]{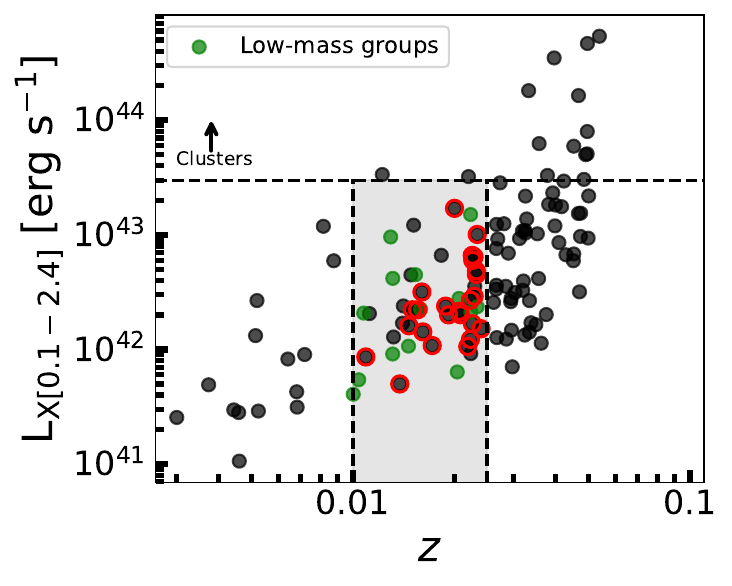}
    \end{subfigure}
    \caption{Selection criteria for the sample studied in this work. \textit{Left:} Total mass -- redshift plane. The parent 2MRS-eRASS1 catalogue with $N_{\rm gal} \geq 5$ and ${\rm nH}\leq7\times10^{20}{\rm cm}^{-2}$ is plotted with black circles. High-$L_{\rm X}$ clusters are shown in green, and our selected 2MRS-eRASS1 sample is bordered with red edges. The bottom-right key explains the selection rationale: the source confusion limit and catalogue completeness (down to our $L_{\rm X}$ limit) dictated our low and high redshift limits, respectively. The horizontal line denotes our mass cut at $\geq2\times10^{13}M_{\odot}$, where the fraction of the eRASS1 detections of 2MRS groups becomes lower than $50\%$. \textit{Right:} X-ray luminosity -- redshift plane. Black and red markers have the same meaning, but the green circles now show the low-mass systems defined below the mass cut. The horizontal line represents the group regime limit defined at $\leq3\times10^{43}$erg $\rm{s}^{-1}$.}
    \label{selection}
\end{figure*}
Energetic processes associated with matter accretion onto supermassive black holes, known as active galactic nuclei(AGN) feedback, are currently the favoured scenario for ejecting gas outside the virial regions of galaxy groups and clusters and ultimately reshaping the matter distribution of the Universe on small scales \citep[]{rudd08, semboloni11, fabian12, sommerville15, vandaalen2020, salcido23, sorini22, baryonification2}. The energy released by AGNs can be comparable to or even exceed the gravitational binding energy of group-scale halos, making galaxy groups the ideal laboratories for studying the impact of feedback on baryon distributions (for comprehensive reviews, see \citealt{eckert21} and \citealt{oppen21}). Unlike massive clusters, where deeper potential wells retain most of the baryons, groups retain only a fraction of their expected gas, with the missing baryons thought to reside in the warm-hot intergalactic medium at larger radii or to have been ejected beyond the virial boundary entirely.

The current observational consensus, based on \textit{Chandra}, \textit{XMM-Newton}, and \textit{Suzaku} observations, is that the gas fractions inside $R_{\rm 500}$ of galaxy groups are indeed sub-cosmic \citep{gastaldello07, pratt09, Lovisari15, xxl, lovisari22, suzaku_gas}. At $R_{\rm 200}$, however, measurements become increasingly scarce at group scales due to the limited sensitivity of X-ray instruments and the elevated and poorly constrained background levels at such large radii. The few existing studies suggest gas fractions that either reach or remain slightly below the cosmic value \citep[e.g.][]{suzaku_gas, tholken16}. This uncertainty at large radii is particularly problematic because it is precisely in these outer regions that the signatures of feedback — gas that has been heated and expelled — should be most evident.

Parallel to observational progress, modern cosmological hydrodynamical simulations have made significant strides in modelling the complex interplay between gravity, gas physics, and feedback processes. Simulations such as IllustrisTNG \citep{illustristng}, SIMBA \citep{simba}, EAGLE \citep{eagle}, BAHAMAS \citep{bahamas17}, and FLAMINGO \citep{flamingo} now routinely predict gas fractions out to several $R_{\rm 500}$, though their predictions exhibit notable discrepancies traceable to different AGN feedback implementations. Due to resolution limitations, feedback is modelled using subgrid parameters typically tuned to match observed gas fractions at $R_{\rm 500}$, meaning predictions beyond this radius represent genuine extrapolations not directly constrained by calibration data. Observational constraints on $f_{\rm gas}$ beyond $R_{\rm 500}$ are therefore particularly valuable as critical tests of a feedback models' fidelity. 

The importance of constraining baryon distributions extends beyond the study of individual halos. Baryonic feedback processes, by ejecting gas from halos, suppress the matter power spectrum (MPS) on scales $k\sim 5-30 h \ \rm{Mpc}^{-1}$, with group-scale halos ($M_{\rm 200} = 10^{13}-10^{14}M_{\odot}$) contributing most significantly \citep{miller25}. This suppression is a key uncertainty for upcoming large-scale structure surveys such as the Legacy Survey of Space and Time \citep{lsst} and those being carried out by \textit{Euclid} \citep{euclid} and the \textit{Nancy Grace Roman} Space Telescope \citep[e.g.][]{roman1}. Moreover, direct baryon budget measurements in groups provide essential input for baryonification correction schemes \citep{schneider15, schneider19, giri21, Debackere20}.

The extended ROentgen Survey with an Imaging Telescope Array (eROSITA) aboard the Spectrum-Roentgen-Gamma mission \citep{eROSITA} offers a transformative capability for addressing these questions. Its large field of view (FOV), stable background, and excellent point spread function (PSF) are ideal for detecting low-surface brightness (SBx) emission beyond the virial radii of nearby groups \citep[][]{merloni24, erass1, bahar24}. Recently, we constructed the AXES-2MRS catalogue \citep{axes2mrs} by cross-matching ROSAT megaparsec-scale X-ray sources with Two Micron Redshift Survey (2MRS) optical groups \citep{tempel18}. A reanalysis of public eROSITA data \citep{f_inprep} confirmed these sources and identified new counterparts. The high detection fraction ($>50\%$ for massive 2MRS systems) provides a complete, unstacked view of X-ray emission from optically selected groups that preserves system-to-system variations.

We exploited these advantages to investigate the hot gas mass fractions at $R_{\rm 500}$ and $R_{\rm 200}$ for a well-defined sample of 25 galaxy groups, and to construct scaling relations between X-ray observables and total mass. Our study was performed at low redshifts ($0.01 < z < 0.025$), ensuring that the scales of interest are well resolved (the typical $R_{\rm 500c}$ of our groups is $\sim 20$ arcmin) and that contamination from unresolved galactic emission can be fully excluded. We compared these observations with hydrodynamical simulations to assess AGN feedback implementations and constrain baryon distribution physics.

In Sect. \ref{sectiondata} we present the sample selection and describe the 2MRS and eROSITA datasets. In Sect. \ref{section: analysis} we detail our analysis of the eROSITA X-ray data, including SBx profile extraction and the derivation of gas mass and gas fraction profiles. In Sect. \ref{section:results} we present our results on gas fractions and scaling relations, including a careful treatment of aperture covariance. We discuss the implications of our findings in the form of constraints on the matter power spectrum in Sect. \ref{sec:mps} and summarise our findings in Sect. \ref{sec:summary}. Throughout this study, we adopt a flat $\Lambda$ cold dark matter ($\Lambda$CDM) cosmology with parameters $H_{0} = 70$ km s$^{-1}$ Mpc$^{-1}$, $\Omega_{\rm m} = 0.3$, and $\Omega_{\Lambda} = 0.7$. Unless otherwise stated, errors represent $1\sigma$ uncertainties (68\% confidence level).

\section{Sample}
\label{sectiondata}
 The complete cross-correlation between the large-scale X-ray sources in the first public release of eROSITA-DE data \citep[eRASS1;][]{merloni24} and the optical galaxy clustering in 2MRS \citep[][]{tempel18} is the subject of an accompanying paper \citep{f_inprep}. In this work, we focused on a subsample defined by the redshift range $0.01 < z < 0.025$ and describe it below.

\subsection{2MRS dataset}
The 2MRS group catalogue is an optical galaxy group sample constructed from the 2MASS Redshift Survey galaxy sample \citep{2012ApJS..199...26H}, with galaxies brighter than $K_{\rm S} = 11.75$ mag,  and is highly complete above Galactic latitudes of $|b| > 5^{\circ}$. It employs a probabilistic approach, modelling the groups within a Bayesian framework using a marked point process model. In practice, this probabilistic algorithm produces groups very similar to those identified by the widely used Friends of Friends algorithm \citep{Tempel16}. The use of 2MRS in the identification of large X-ray sources has been adopted recently in \citet{axes2mrs} where we presented all ROSAT All-Sky Survey (RASS) sources that have an optical galaxy group counterpart at $z<0.07$ down to three spectroscopic members.

\subsection{eRASS1 catalogue}
\label{sub:erass1}
We performed our analysis with the released eRASS1 event lists that cover the Galactic longitudes $l > 179.6^{\circ}$. Due to the `light leak' problem in two of the seven eROSITA telescope modules (TM5 and TM7), where optical solar photons reach the detectors creating fake X-ray events with energies $\leq0.5$ keV \citep{eROSITA} --- as well as the large-scale soft X-ray filament emission --- we worked with [$0.6-2.3$] keV images and corresponding exposure maps to eliminate these effects from our analysis. We followed the same extended source detection method and determination of the flux extraction regions applied in, for example, \citet{kafer19}, \citet{Finoguenov20}, and \citet{axes2mrs}. Briefly, a multi-scale wavelet decomposition method\footnote{\url {https://github.com/avikhlinin/wvdecomp}} \cite[][]{vikhlinin98} is used where an image reconstruction is made for each of $4\times2^{n-1}$ arcsecond scales, with $n\in [2,9]$. We show the wavelet images describing the X-ray emission of our group sample in Appendix \ref{wv_gallery}. To avoid missing any sources, we identified structures on separate scales and merged them afterwards. We used \texttt{SExtractor} \citep{sextractor} to produce catalogues of point sources using the spatial scales below one arcminute, which were visually inspected and then masked in the subsequent analysis. The sensitivity to point sources is significantly higher for eROSITA than for RASS, primarily due to eROSITA’s substantially improved PSF. This creates a large separation between the flux detection limits for point-like and extended emission, computed as
\begin{eqnarray}
    \rm factor = \frac{\rm Flux \ of \ extended \ source}{Limit \ of \ point \ source \ detection},
\end{eqnarray}
and it is typically a factor of 30 for our eROSITA study. This means that to explain the unresolved X-ray emission, one would require an excess of hundreds of undetected point sources, exceeding the number of group members by an order of magnitude.

\subsection{2MRS-eRASS1}

We used dynamical masses from the 2MRS group catalogue, limited to groups with at least five members where these estimates are reliable, and verified their consistency with the AGN feedback simulations of \citet{munari13}. Total masses ($M_{\rm 500c}$ and $M_{\rm 200c}$) and $R_{\rm 500c}$ are listed in Tables \ref{table: data} and \ref{table: mass}. X-ray emitting groups are selected via the modified Hausdorff distance method of \citet{kosowski25}, as described in \citet{f_inprep}, and in this work we studied their X-ray emission and constrain the hot gas budget out to $R_{\rm 200c}$ with eROSITA.

\begin{figure}
\centering
   \includegraphics[width=0.47\textwidth]{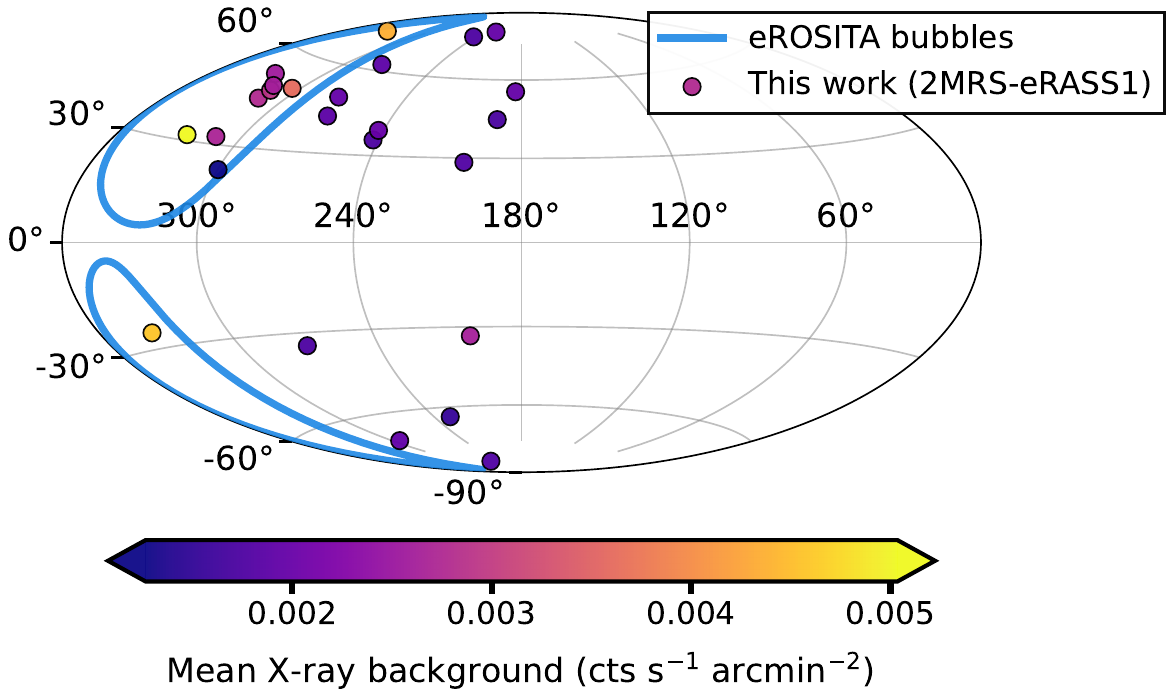}
     \caption{All-sky distribution of the 2MRS-eRASS1 galaxy group sample with respect to the eROSITA X-ray bubbles, using Galactic coordinates and a Hammer-Aitoff projection. The elliptical boundaries represent the approximate extent of the bubbles at $ l \gtrsim 290^{\circ}$ in the north and $ l \gtrsim 320^{\circ}$ in the south \citep{Yeung24}. Groups are colour-coded by their mean X-ray background.}
     \label{bubble}
\end{figure}

Our redshift range $0.01 < z < 0.025$ is conservatively chosen to ensure sensitivity to group outskirts within eRASS1 spatial scales and to avoid source confusion, and is further guided by the AXES-2MRS completeness limit down to $L_{\rm X} > 3\times10^{42}$ erg s$^{-1}$ (Figs. 2 and 10 in \citealt{axes2mrs}). We further imposed $L_{\rm X} < 3\times10^{43}$ erg s$^{-1}$ to remain in the group regime \citep{vulcani26}, applied the typical neutral hydrogen column density extragalactic definition of $\rm nH < 7\times10^{20}$ cm$^{-2}$, and adopted a mass limit of $M_{\rm 500c} > 2\times10^{13}M_{\odot}$, below which the fraction of X-ray sources identified as 2MRS groups drops below $50\%$ \citep{f_inprep}. The full selection is illustrated in Fig. \ref{selection}, and the group locations, including the eROSITA X-ray bubbles \citep{bubble}, are shown in Fig. \ref{bubble}.

Our selection criteria initially yielded a sample of 37 groups. However, 12 systems were excluded: eight due to elevated X-ray background exceeding twice the quiescent level within $0.5^{\circ}$; two eROSITA tiles (192099 and 192105) affected by projection effects from overlapping optical systems with 3D separations below 100 Mpc; one field (303132) where most X-ray emission falls outside German proprietary rights; and one system (194102) at the FOV boundary with only a detectable core. The remaining 25 groups define our 2MRS-eRASS1 sample. 

\section{Data analysis}
\label{data_analysis}

We used the standard eROSITA Science Analysis Software System \citep[\texttt{eSASS4DR1};][]{esass} version 010 as well as its calibration database (\texttt{CALDB}) to process the event lists into calibrated ($3.6 \times 3.6$ deg$^{2}$) count and exposure maps. In more detail, we used the \texttt{evtool} task to create images and filtered them with \texttt{flaregti} to remove flared times, \texttt{PATTERN $\leq 15$} to keep all valid events, and \texttt{FLAG=0xc00fff30} to exclude bad pixels and their neighbouring ones. Finally, we excised the $[1.4-1.6]$ keV energy band associated with the Al-K$\alpha$ instrumental line, as its fluorescence was found to be consistently $\sim20-60\%$ lower in the filter-wheel closed (FWC) data for all telescope modules (TMs; \citealt{Yeung24, perinati25}). We still refer to our used band as $[0.6-2.3]$ keV for simplicity. We produced exposure maps with similar settings using the \texttt{expmap} task. 

Particle-induced background (PIB) maps were generated for each sky tile by measuring count rates in the hard [6--9] keV band and scaling to the [0.6--2.3] keV band of interest using a fixed ratio derived from FWC data, following the procedure of \citet{pib}. The resulting maps are spatially distributed using unvignetted exposure maps. Full details of the procedure are provided in Appendix~\ref{section:pib}. The eROSITA PSF model was calibrated in-flight using three point sources from eRASS1, whose radial SBx profiles were fitted with a King model (Eq.~\ref{eq:beta}). The best-fit parameters are $R_{\rm c} = 6$ arcsec and $\alpha=1.19$, and the model was validated by recovering exactly 50\% of the total flux within the eROSITA half-energy width of 30 arcsec. The details of this process are provided in Appendix~\ref{sec:psf}.

\label{section: analysis}

\begin{figure}
\centering
   \includegraphics[width=0.47\textwidth]{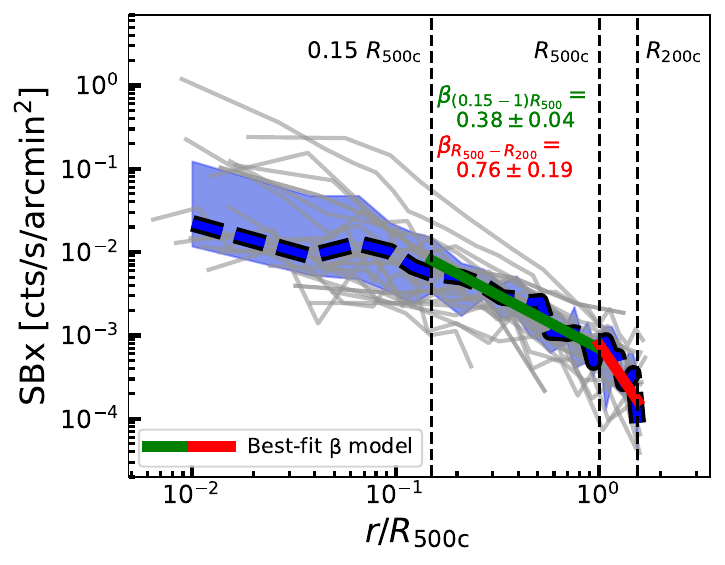}
     \caption{Background-subtracted X-ray SBx profiles ([$0.6-2.3$]~keV) of the 2MRS-eRASS1 group sample. Vertical dashed lines mark $0.15R_{500\rm c}$, $R_{500\rm c}$, and $R_{200\rm c}$; the thick dashed blue curve is the median profile. Green and red curves show the best-fit $\beta$~models of the median profile in the $[0.15-1]R_{500\rm c}$ and $R_{500\rm c}-R_{200\rm c}$ ranges, respectively. The stacked profile steepens beyond $R_{500\rm c}$, with $\beta$ rising from $0.38\pm0.04$ to $0.76\pm0.19$.} 
     \label{profiles}
\end{figure}

\subsection{Surface brightness profiles and luminosities}
\label{section: sb}
Surface brightness profiles were extracted from the [$0.6-2.3$] keV eROSITA images from all seven TMs (TM0) in logarithmic radial bins centred on the large-scale X-ray peak, out to $R_{200\rm c}$, with a minimum bin size of 30~arcsec, using \texttt{pyproffit} \citep{pyproffit}, accounting for exposure map correction, particle and X-ray background subtraction, and PSF modelling. The X-ray background was extracted from multiple regions across the FOV of each group and subtracted from the source regions (the procedure is described in Appendix \ref{sec:bkg}). Nine systems showed an elevated and spatially varying background; these lie within the eROSITA Galactic foreground bubble pair \citep{bubble}, which extends to $|b|\approx85^{\circ}$ at $l\gtrsim290^{\circ}$ (north) and $l\gtrsim320^{\circ}$ (south; \citealt{Yeung24}; sample location relative to the bubbles shown in Fig.~\ref{bubble}). Excluding these systems has no significant impact on the scaling relations beyond a marginal flattening of the $M_{\rm gas}-M_{\rm tot}$ relation, confirming the robustness of our multi-region background subtraction.

The central $<0.15R_{500\rm c}$ region was excised when fitting SBx profiles and computing X-ray luminosity, since we found the scatter in the $L_{\rm X,500}-M_{\rm gas,500}$ relation increases by a factor of $1.8\pm0.4$ using the full-range $L_{\rm X}$, consistent with other studies where core inclusion enhances scatter by at least a factor of 2 \citep[e.g.][]{lovisari22, pratt09}. We computed core-excised luminosity in the standard $[0.1-2.4]$~keV band for two apertures, $[0.15-1]R_{500\rm c}$ and $0.15R_{500\rm c}-R_{200\rm c}$, and list them  in Table~\ref{table: data}.

We fitted the profiles with a 1D beta model \citep{beta_model} that has the same form as the King model (Eq. \ref{eq:beta}) with $\alpha = 3\beta-0.5$, where $\boldsymbol{\alpha}$
represents the slope at large radii ($r>>R_{\rm c}$). To investigate profile steepening beyond $R_{\rm 500c}$, a high S/N is required to obtain reliable constraints in the outskirts. We therefore stacked the profiles of the sample into 15 logarithmically spaced bins spanning $0.15R_{\rm 500c}-R_{\rm 200c}$, and fitted the resulting profile with a beta model. The resulting profiles and best-fit parameters are presented in Sect. \ref{sub:beta_params}.

\begin{figure*}[hbt!]
    \centering
    \begin{subfigure}[b]{0.45\textwidth}
        \hspace{-0.1in}
            \includegraphics[height=0.28\textheight]{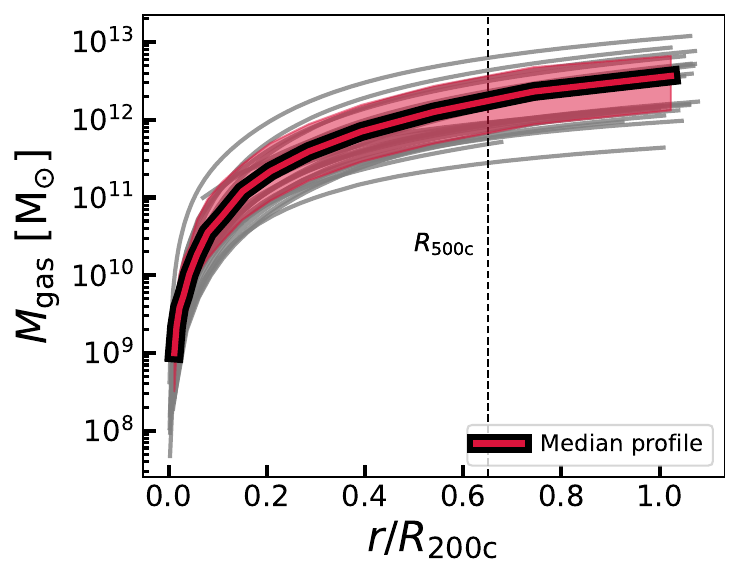}
    \end{subfigure}
    \begin{subfigure}[b]{0.45\textwidth}
            \hspace{0.1in}
            \includegraphics[height=0.28\textheight]{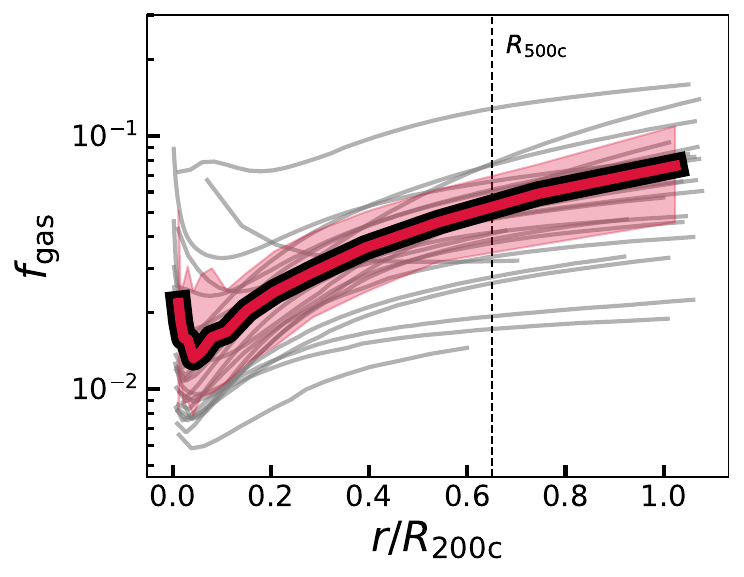}
    \end{subfigure}
    \caption{Hot gas mass and hot gas fraction profiles of the 2MRS-eRASS1 groups. \textit{Left:} Hot gas mass profiles in units of $R_{\rm 200c}$. The vertical dashed black line marks $R_{\rm 500c}$. \textit{Right:} Hot gas fraction profiles in units of $R_{\rm 200c}$.}
    \label{mgas_fgas_prfs}
\end{figure*}

\begin{figure}
\centering
   \includegraphics[width=0.47\textwidth]{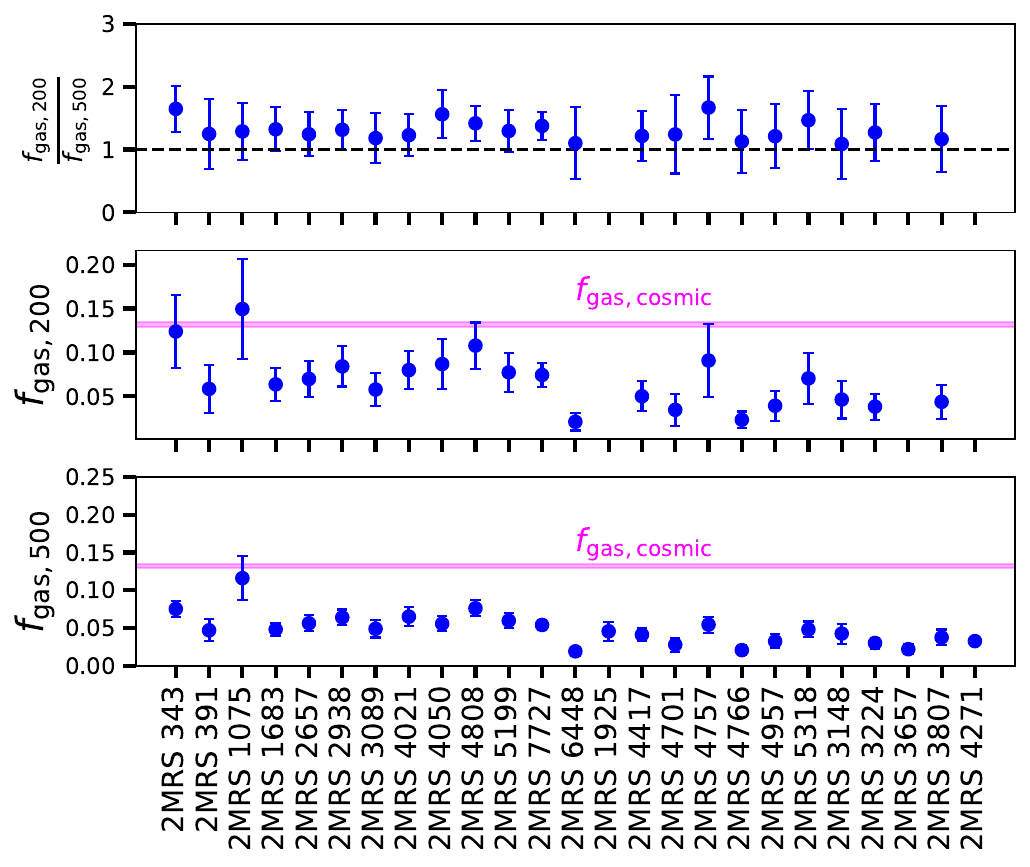}
     \caption{Cumulative hot gas mass fractions of the 2MRS-eRASS1 sample. \textit{Bottom:} $f_{\rm gas}$ inside $R_{\rm 500c}$. 
     \textit{Middle:} $f_{\rm gas}$ inside $R_{\rm 200c}$. \textit{Top:} Ratio between $f_{\rm gas}$ inside $R_{\rm 200c}$ to that inside $R_{\rm 500c}$, with the dashed~black line marking $f_{\rm gas,200}=f_{\rm gas,500}$.} 
     \label{fgas_all}
\end{figure}

\subsection{Gas mass}
\label{analysis: fgas}
Following \citet{xxl}, we used \texttt{pyproffit}\footnote{\url{https://pyproffit.readthedocs.io/en/latest/}} \citep{pyproffit} to convert SBx into gas mass profiles. First, we simulated an absorbed APEC model \citep{apec} to convert SBx profiles to emission measure (EM) profiles, assuming a constant temperature from the $L_{\rm X}-T$ relation of the XXL-100-GC bright cluster sample \citep{giles16}:
\begin{eqnarray}
   \frac{L_{\rm X}}{3 \times 10^{43} \text{ erg s}^{-1}} = E(z)^{1.64} \times 0.71 \times \left(\frac{T}{3 \text{ keV}}\right)^{2.63}.
\end{eqnarray}
The implied median $kT$ of our sample is 1.2 keV ([$20^{\rm th}-80^{\rm th}$] percentiles: [0.88 - 1.54 keV]). The XXL relation intrinsic scatter of $L_{\rm X}$ is $47\%$, which translates to $18\%$ on $kT$ ($\sigma_{kT}=\sigma_{L_{\rm X}/\rm slope}$). This yields an $8\pm2\%$ systematic uncertainty on our measured gas mass. {We also remark that 
for systems with $kT \sim 1$ keV, adopting a single temperature and metallicity to characterise each system can introduce a systematic uncertainty of up to $\sim30\%$ in the inferred gas mass. \citep{eckert25}.

The Galactic absorption value for each group was adopted from \citet{hi4pi}'s neutral hydrogen column density maps (\texttt{nH}\footnote{\url{https://heasarc.gsfc.nasa.gov/cgi-bin/Tools/w3nh/w3nh.pl}}). The metallicity was fixed to 0.3 $Z_{\odot}$ based on \citet{anders89} solar abundances. This allowed us to compute a conversion factor between the EM and the APEC normalisation parameter that is related to the EM according to
\begin{eqnarray}
   \rm Norm_{\rm APEC} = \frac{10^{-14}}{4\pi[D_{\rm A}(1+z)]^2} \int n_{\rm e}n_{\rm H} \mathrm{d}V,
\end{eqnarray}
where $D_{\rm A}$ is the angular diameter distance to the source. In this step, we used the exposure and PSF-corrected SBx profiles, properly rescaled using the on-axis effective area of eROSITA. 

We assumed the gas density is the sum of the electron and proton densities in the fully ionised plasma $\rho_{\rm gas} = n_{\rm e}+n_{\rm H}$ with $n_{e} = 1.17 n_{\rm H}$. Next, a multi-scale forward-fitting method was used to deproject the observed 2D EM profiles into 3D density profiles by decomposing them into a sum of King functions, each deprojected individually assuming spherical symmetry. We refer the reader to \citet{pyproffit}, Appendix B of \citet{xxl}, and Appendix D of \citet{seppi25} for a detailed description of the method. In extracting the SBx profiles, we employed fine radial log binning with a minimum bin size between 10 and 30 arcsec adaptively chosen to ensure a reliable S/N in each annulus. 

Using a mean molecular weight of $\mu = 0.61$ and the proton mass $m_{\rm p}$, the cumulative gas mass within a radius $r$ was then obtained by integrating the enclosed density profile 
\begin{eqnarray}
   M_{\rm gas}(<r) = \mu m_{\rm p}\int_0^r 4\pi r^{2} n_{\rm e}(r)\ dr. 
\end{eqnarray}

\subsection{Total mass}
We modelled the total mass using a Navarro-Frenk-White (NFW) profile \citep{nfw} with \citet{duffy08} concentrations ($c_{\rm 200c}$):
\begin{eqnarray}
   M_{\rm NFW}(<r) = 4\pi  \rho_{\rm s}  r_{\rm s}^{3}  \left[\ln{\left(1+\frac{r}{r_{\rm s}}\right)} - \frac{r}{r + r_{\rm s}}\right], 
\end{eqnarray}
where the NFW scale radius $r_{\rm s} = r_{\rm 200c}/c_{\rm 200c}$ and the characteristic density $\rho_{\rm s} = M_{\rm 200c}/[4\pi r_{\rm s}^{3} f(c)]$ with $f(c) = \ln{(1+c)}-c/(1+c)$. The optical $M_{\rm 200c}$ values from 2MRS agree within $5\%$ of the standard \citet{munari13} velocity dispersion-mass relation.

\subsection{Gas fraction}
To marginalise over the concentration parameter distribution from \citet{duffy08} and correctly propagate all of the uncertainties associated with $M_{\rm 200c}$, $c_{\rm 200c}$, and $M_{\rm gas}$ to the gas mass fraction profile calculation, a Monte Carlo simulation was constructed where for each iteration:

\begin{itemize}
    \item An $M_{\rm 200c}$ value was drawn from a log-normal distribution parametrised using the mean $M_{\rm 200c}$ and its $1\sigma$ deviation ($M_{\rm 200c,i}$).
    \item $M_{\rm 200c,i}$ was then used to obtain a mean $c_{\rm 200c}$ from \citet{duffy08}'s concentration-mass relation ($c_{\rm mean,i}$).
    \item A concentration value was drawn from a log-normal distribution centred on $c_{\rm mean,i}$ with an intrinsic scatter of that reported in the relation $\sigma_{\log_{10}} = 0.15$ dex ($c_{\rm i}$).
    \item Hot gas mass values at each radius were drawn from log-normal distributions parametrised using the mean cumulative gas mass and their $1\sigma$ deviation ($M_{\rm gas,i}(<r)$).
    \item The hot gas mass fraction profile was calculated for this iteration as $f_{\rm gas,i}(<r) = \frac{M_{\rm gas,i}(<r)}{M_{\rm NFW,i}(<r)}$.
\end{itemize}
The adopted $f_{\rm gas}$ profile is the median of $f_{\rm gas,i}(<r)$ across 10,000 iterations. The 16th-84th percentiles were calculated and now reflect the propagated uncertainties from $M_{\rm 200c}$, $c_{\rm 200c}$, and $M_{\rm gas}$. We note that one system (2MRS: 6696, eROSITA tile: 303132) was on the edge of the eROSITA-DE footprint, and thus we could not extract the profile out to $R_{\rm 500c}$. Another system (2MRS: 4401, eROSITA tile: 194102) was on the edge of the FOV, so we could only reach $0.75 R_{\rm 500c}$. Both systems were omitted from the gas fraction analysis.

\section{Results}
\label{section:results}

\subsection{Beta model parameters}
\label{sub:beta_params}

The particle and X-ray sky background-subtracted SBx profiles of 2MRS-eRASS1 are shown in Fig. \ref{profiles}. We detect uniformly flat $\beta$ parameter values across our sample with a best-fit $\beta$ value of $0.38 \pm 0.04$ when fitting the stacked profile in the range $[0.15-1]R_{\rm 500c}$. We report the individual $\beta$ values in Table \ref{table: beta}. Remarkably, we detect steepening of the density profiles of our group sample, indicated by the best-fit $\beta$ value for the stacked profile outside $R_{\rm 500c}$ of $\beta_{0.15  R_{\rm 500c}-R_{\rm 200c}} = 0.76 \pm 0.19$. This steepening suggests that the relative ejection of baryons beyond $R_{\rm 200c}$ weakens, providing a first point of tension with strong feedback models, which preferentially suppress core densities while leaving the outskirts close to fiducial values, producing flatter rather than steeper outer profiles (see Fig. 7 in \citealt{bras24}). A similar trend between the steepening of the profiles and the AGN feedback strength in galaxy groups is also seen in X-GAP-like \citep{xgap} groups in FLAMINGO (Seppi et al. in prep.). We explore this tension further in Sects. \ref{sec:lxm} and \ref{mgas-m}.

The core radius of the beta model is a difficult measurement in the galaxy group regime. The SBx profiles in our sample rise like power laws towards the central regions with no well-defined cores. We report extremely small core radii of less than $0.1'$ and present our best-fit $R_{\rm C}$ constraints in Table \ref{table: beta}. We remark that our $R_{\rm C}$ values are sensitive to the detailed shape of the PSF model. Our results are in agreement with the recent study of \citet{Li24} who stacked [0.2-2.3] keV X-ray images of a large optically selected sample of poor galaxy groups (defined as having fewer than five members) based on the Dark Energy Spectroscopic Instrument (DESI) Legacy Imaging Surveys and cross-matched with the eROSITA Final Equatorial-Depth Survey (eFEDS). They divide their sample into 16 subsamples based on redshift and group mass, and report a best-fit $\beta$ value of $0.41^{+0.07}_{-0.04}$ in the bin most closely comparable to our sample ($0.1 < z< 0.2$ and $13.0 < \log{M}/M_{\odot} < 13.5$).

\subsection{Hot gas fraction}
The cumulative hot gas fraction is defined as the median of $f_{\rm gas,i}(<r)$ across 10,000 Monte Carlo iterations (see Sect. \ref{analysis: fgas} for details). We present the hot gas mass and the hot gas mass fraction radial profiles of 2MRS-eRASS1 in Fig. \ref{mgas_fgas_prfs}, and show the resulting $f_{\rm gas,500}$ and $f_{\rm gas,200}$ values in the bottom and middle panels of Fig. \ref{fgas_all}, respectively. We show the diagnostic quantity $\frac{f_{\rm gas}(<R_{\rm 200c})}{f_{\rm gas}(<R_{\rm 500c})}$, which is the ratio between the gas fraction at $R_{\rm 200c}$ to that at $R_{\rm 500c}$, in the upper panel of Fig. \ref{fgas_all}. This parameter quantifies the baryon redistribution outside the virial regions by AGN feedback, and serves as a direct constraining parameter across the predictions of different numerical simulations. The cosmic gas mass fraction marked in the figure at $f_{\rm gas, cosmic} = 0.132 \pm 0.003$, is adopted based on the cosmic baryon fraction value $f_{\rm b,cosmic} = 0.157 \pm 0.003$ and a stellar mass fraction of $f_{\rm *} = 0.025$ based on the IllustrisTNG100 prediction for groups with total mass $M_{\rm 200c} \sim 10^{13.5}M_{\odot}$ (see the middle panel of Fig. 3 in \citealt{oppen21}).

At our median mass ($M_{\rm 500c} = 2.54\times10^{13} M_{\odot}$), we obtain $f_{\rm gas,500} = 4.32 \pm 0.42\%$, in agreement at a $1.6\sigma$ level with \citet{popesso_gas24}'s eRASS1 stacking results for the Galaxy And Mass Assembly survey (GAMA) groups in the eFEDS area spanning $10^{12}M_{\odot}-10^{14} M_{\odot}$ and $z < 0.2$. At $R_{\rm 200c}$ and median mass ($M_{\rm 200c} = 3.69\times 10^{13}M_{\odot}$), we find $f_{\rm gas, 200} = 5.78 \pm 0.69\%$, fully consistent with \citet{popesso_gas24}'s $f_{\rm gas,200}$ at the same mass (see Sect. \ref{mgas-m}). We also compared our eROSITA gas masses with published results for the seven groups in our sample for which published \textit{Chandra} studies are available (see Appendix \ref{sec:comp}).

\subsection{Scaling  relations}
\label{sec: scaling relations}

\begin{table*}[hbt!]
\caption{Fitting formulas and best-fit parameters of the scaling relations defined in Eq. \ref{scaling_eqn}.}
\label{tab: relations}
\centering
\begin{tabular}{c c c c c c}
\hline\hline
Relation ($Y-X$) & $A$ & $B$ & $\sigma_{\mathrm{int}}$ & $C_{1}$ & $C_{2}$\\
\hline
$M_{\rm gas, 500}-M_{\rm 500c}$ & $0.15^{+0.11}_{-0.12}$ & $1.59^{+0.34}_{-0.28}$ & $0.23^{+0.12}_{0.11}$ & $8.96\times10^{11}M_{\odot}$ & $2.54\times10^{13}M_{\odot}$\\
$M_{\rm gas, 200}-M_{\rm 200c}$ & $-0.16^{+0.12}_{-0.13}$ & $1.53^{+0.26}_{-0.24}$ & $0.28^{+0.14}_{-0.13}$ & $2.52\times10^{12}M_{\odot}$ & $3.69\times10^{13}M_{\odot}$\\
$L_{\rm X, 500}-M_{\rm gas, 500}$ & $0.03\pm0.07$ & $1.04\pm0.07$ & $0.16\pm0.07$& $1.32\times10^{42}$erg s$^{-1}$ & $8.96\times10^{11}M_{\odot}$\\
$L_{\rm X, 200}-M_{\rm gas, 200}$ & $0.09 \pm 0.06$ & $1.02 \pm 0.08$ & $0.20^{0.05}_{0.04}$ & $2.41\times10^{42}$erg s$^{-1}$ & $2.52\times10^{12}M_{\odot}$\\
$L_{\rm X,500}-M_{\rm 500c}$ & $0.13^{+0.16}_{-0.18}$ & $1.78^{+0.54}_{-0.45}$ & $0.41\pm0.17$ & $1.32\times10^{42}$erg s$^{-1}$ & $2.54\times10^{13}M_{\odot}$\\
$L_{\rm X,200}-M_{\rm 200c}$ & $-0.12^{+0.15}_{-0.16}$ & $1.55^{+0.25}_{-0.28}$ & $0.28^{+0.19}_{-0.15}$ & $2.41\times10^{42}$erg s$^{-1}$ & $3.69\times10^{13}M_{\odot}$\\

\hline
\end{tabular}

\end{table*}

We fitted a power-law to the scaling relation between the observables $[L_{\rm X}, M_{\rm tot}, M_{\rm gas}]$ of the form
\begin{eqnarray}
    \ln{\left(\frac{Y}{C_{1}}\right)} = A + B \ln{\left(\frac{X}{C_{2}}\right)} + \sigma_{\mathrm{int}},
    \label{scaling_eqn}
\end{eqnarray}
where $A$, $B$, and $\sigma_{\mathrm{int}}$ are the normalisation, slope, and intrinsic scatter of the relations, respectively. $\sigma_{\mathrm{int}}$ is measured in the form $\sigma_{\ln{Y}|\ln{X}}$. The medians and/or pivots $C_{1}$ and $C_{2}$ for each relation are listed in Table \ref{tab: relations}. 

\subsubsection{Aperture covariance correction}
\label{sec:aper}
The integrated quantities $M_{\rm tot}(<r)$ and $M_{\rm gas}(<r)$ are covariant because they depend on the same radius of integration $r$. In other words, the shared dependence of $M_{\rm 500c}$ ($M_{\rm 200c}$) and $M_{\rm gas,500}$ ($M_{\rm gas,200}$) on $R_{\rm 500c}$ ($R_{\rm 200c}$) suppresses the physical variation of $M_{\rm gas}$ (and thus $f_{\rm gas}$) with the total mass, leading to artificially flatter $f_{\rm gas}-M_{\rm tot}$ relations. When $M_{\rm 500c}$ is up-scattered, $R_{\rm 500c}$ is also up-scattered and $M_{\rm gas}$ increases accordingly. We can quantify this effect as follows: the electron density of an isothermal sphere of radius $r$ can be modelled as \citep{rho_e}

\begin{eqnarray}
    \rho_{\rm gas}(r) = \rho_{0} \left[ 1+ \left(\frac{r}{R_{C}} \right)^{2} \right]^{-3\beta/2},
    \label{density}
\end{eqnarray}
where $\beta$ is the same parameter studied in Sect. \ref{section: sb}. At $r>>R_{C}$, $\rho_{\rm gas}\propto r^{-3\beta}$ and since $M_{\rm gas} = 4\pi\int{dr \rho_{\rm gas} r^{2}}$, then
\begin{eqnarray}
    M_{\rm gas}\propto M_{\rm tot}^{1-\beta} \longrightarrow f_{\rm gas}\propto M_{\rm tot}^{-\beta},
\end{eqnarray}
  which is exactly the geometric effect of decreasing $f_{\rm gas}$ with total mass. Similarly, one can write
   \begin{eqnarray}
    L_{\rm X} \propto M_{\rm tot}^{1-2\beta}.
\end{eqnarray}
Assuming that the observed $\hat{M}^{\rm obs}_{\rm gas}$, $\hat{L}^{\rm obs}_{\rm X}$, and $\hat{M}^{\rm obs}_{\rm tot}$ are noisy realisations of the covariant quantities $\hat{M}_{\rm gas}$, $\hat{L}_{\rm X}$, and $\hat{M}_{\rm tot}$, which are related to the true underlying $M_{\rm gas}$, $L_{\rm X}$, and $M_{\rm tot}$ according to
\begin{eqnarray}
   \hat{M}_{\rm gas} = M_{\rm gas} \left(\frac{\hat{M}_{\rm tot}}{M_{\rm tot}}\right)^{1-\beta} \quad \text{and}  \quad \hat{L}_{\rm X} = L_{\rm X} \left(\frac{\hat{M}_{\rm tot}}{M_{\rm tot}}\right)^{1-2\beta}.
   \label{eq:cov}
\end{eqnarray}
We applied these corrections and present the details of constructing the scaling relations in the following sections. We used the individual best-fit $\beta$ values obtained from fitting the SB$_{\rm X}$ profiles at $R_{\rm 500c}$ for relations at $R_{\rm 200c}$, as individual $\beta$ measurements are only available only at $R_{\rm 500c}$. We tested the impact of using the steeper $\beta$ from the stacked profile at $R_{\rm 200c}$ (Fig. \ref{profiles}) and found no significant difference in the median best-fit parameters.

\subsubsection{Likelihood}
Our hierarchical formalism of the likelihood of the $\hat{M}_{\rm gas}-M_{\rm tot}$ relation can be written as

\begin{align}
   P(\hat{M}_{\rm tot}) =P(\hat{M}_{\rm tot}|M_{\rm tot}) &P(M_{\rm tot})
\end{align}
\begin{align}
\begin{split}
   P(\hat{M}_{\rm gas}|M_{\rm tot},\beta,\theta) =&
   P(\hat{M}_{\rm gas}|M_{\rm gas},\beta,\hat{M}_{\rm tot},M_{\rm tot})\\
   & P(M_{\rm gas}|M_{\rm tot},\theta)P(\hat{M}_{\rm tot}|M_{\rm tot})\\ &P(M_{\rm tot})P(\beta),
\end{split}
\end{align}
where $\theta$ denote the scaling parameters $[A, B, S_{\ln{Y}|\ln{X}}]$, and the first term on the right-hand side, $P(\hat{M}_{\rm gas}|M_{\rm gas},\beta,\hat{M}_{\rm tot},M_{\rm tot})$, is the covariance correction term as described in Eq. \ref{eq:cov}. In Eq. \ref{scaling_eqn}, we model the parameter $Y$ as log-normally distributed around $X$, then the scaling relation term $P(M_{\rm gas}|M_{\rm tot},\theta)$ becomes
\begin{align}
\begin{split}
P(M_{\rm gas}|&M_{\rm tot},\theta) = \\
&\mathcal{L}\mathcal{N}\left(\mu = M_{\rm gas,piv}\ A \ \left(\frac{M_{\rm tot}}{M_{\rm tot,piv}}\right)^{B}, \sigma = S_{\ln{Y}|\ln{X}} \right).
\end{split}
\end{align}
The two terms $P(\hat{M}_{\rm gas}|M_{\rm gas},\beta,\hat{M}_{\rm tot},M_{\rm tot})$  and $P(\hat{M}_{\rm tot}|M_{\rm tot})$ represent the probability of observing $\hat{M}_{\rm gas}$ and $\hat{M}_{\rm tot}$ given the true $M_{\rm gas}$ and $M_{\rm tot}$, respectively. Finally, $P(M_{\rm tot})$ and $P(\beta)$ are the distributions of the true mass and the $\beta$ parameter of our sample, respectively. 

\begin{figure}[hbt!]
\centering
   \includegraphics[width=0.47\textwidth]{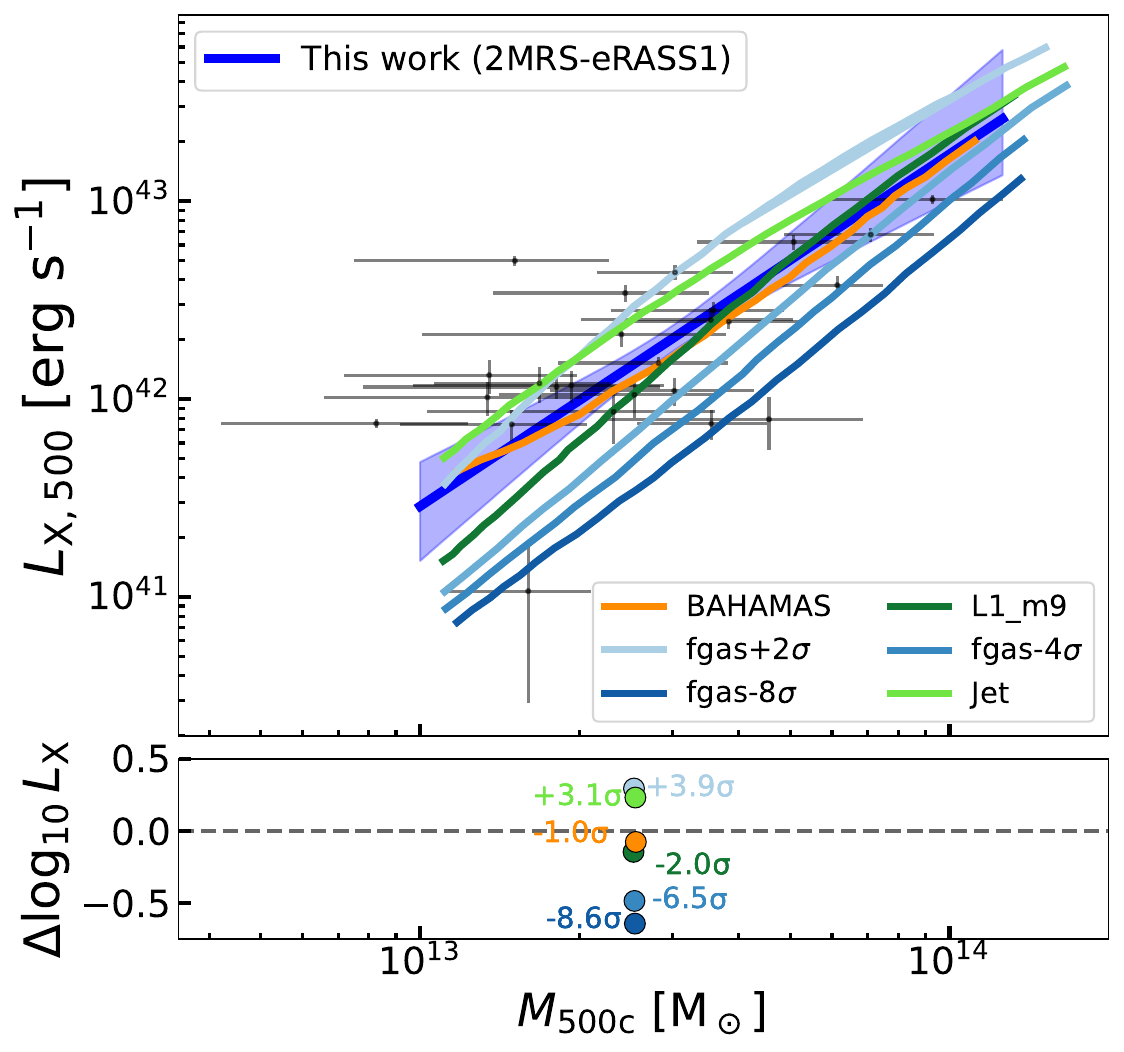}
     \caption{X-ray luminosity vs total mass at $R_{\rm 500c}$. \textit{Top:} Aperture covariance-corrected $L_{\rm X}-M_{\rm tot}$ relation (blue curve) for the 2MRS-eRASS1 sample. The blue shaded region is the $1\sigma$ confidence interval of the slope and normalisation. The orange curve shows the prediction of BAHAMAS, and the other curves represent different FLAMINGO feedback model variations \citep{bras24}. \textit{Bottom:} Difference in $\log_{10}{L_{\rm X}}$ between the explored simulations and our observed relation at the median mass. Text annotations indicate the offset significance at the $\sigma$ level. The relation at $R_{\rm 200c}$ is shown in Fig. \ref{lxm200}.} 
     \label{flamingo_lxm}
\end{figure}

Modelling the distribution of the true underlying mass of our sample is crucial in the hierarchical construction of the scaling relations. The measured total masses we used are dynamical estimates from the 2MRS catalogue, derived from the observed line-of-sight velocity dispersion and the projected distance between galaxy pairs within the groups \citep{tempel14}. Since we did not explicitly model the full selection function of our sample starting from the halo mass function, we needed a physically motivated form for $P(M_{\rm tot})$. To this end, we used the velocity dispersion-true mass calibration by \citet{seppi25}, who established a relation between the true dark matter halo mass in light cones generated from the Uchuu $N$-body simulations \citep{uchuu} and the corresponding measured velocity dispersion of the galaxy members with end-to-end simulations (see their Sect. 6 for more details). This relation provides an unbiased mapping between velocity dispersion and total mass as it takes into account X-ray and optical selection effects.  

In our inference model, we used this relation to get an estimate for the true mass of each group $M_{\rm tot,T}$ given its measured velocity dispersion (see Table \ref{table: data}). We then treated the distribution of true masses in our sample as a latent Gaussian population centred on the mean of $M_{\rm tot,T}$ and with a width that is itself a free parameter. In practice, this corresponds to a prior of the form
   \begin{eqnarray}
    M_{\rm tot} \sim \mathcal{N}(\mu_{\rm MT},\sigma_{\rm MT}),
    \label{eq:true_dist}
\end{eqnarray}
where $\mu_{\rm MT}$ is a Gaussian prior centred on the mean of $M_{\rm tot,T}$ (with the width set by the intrinsic scatter of the \citealt{seppi25} relation), and $\sigma_{\rm MT}$ is a half-normal hyperparameter that captures the sample-level mass spread. Our $P(M_{\rm tot})$ posterior agrees with the expectation from the \citet{seppi25} modelling according to the Kolmogorov-Smirnov two-sample test. Moreover, the distribution mean ($\mu_{\rm MT}$), which is a hyperparameter we fitted for, is consistent within $<1\%$ statistical error with the mean of the expected distribution. We show the comparison between the expected and the posterior $P(M_{\rm tot})$ in Fig. \ref{posterior}.

The likelihood for the $L_{\rm X}-M_{\rm tot}$ relation is constructed similarly by replacing $M_{\rm gas}$ by $L_{\rm X}$ in the previous formalism and using the $L_{\rm X, cov}$ correction instead (Eq. \ref{eq:cov}). On the other hand, the $L_{\rm X}-M_{\rm gas}$ relation has no covariance correction applied and its likelihood reads
\begin{eqnarray}
\begin{split}
    P(\hat{M}_{\rm gas}^{\rm obs},\hat{L}_{\rm X}^{\rm obs}|M_{\rm gas},L_{\rm X},\theta) =& P(L_{\rm X}|M_{\rm gas},\theta)P(\hat{L}_{\rm X}^{\rm obs}|L_{\rm X})\\
    &P(\hat{M}_{\rm gas}^{\rm obs}|M_{\rm gas})P(M_{\rm gas}),
\end{split}
\end{eqnarray}
where $P(M_{\rm gas})$ is the distribution of the true gas masses of our sample, which is modelled, similar to Eq. \ref{eq:true_dist}, as a normal distribution centred on the mean observed gas mass and allowed a varying width.

We implemented the described Bayesian hierarchical framework for the scaling relations using \texttt{PyMC} \citep{pymc}, which samples the posterior distributions with a Hamiltonian Monte Carlo algorithm. The best-fit median parameters are listed in Table \ref{tab: relations} and their 1D and 2D posterior distributions are shown in Appendix \ref{sec: params}. Sampling is performed using four chains, each with 20,000 draws after 7,000 tuning iterations. We present below the $L_{\rm X}-M_{\rm tot}$ and $M_{\rm gas}-M_{\rm tot}$ relations, and we discuss the $L_{\rm X}-M_{\rm gas}$ in Appendix \ref{sec: lxmg}.

\subsubsection{$L_{\rm X}-M_{\rm tot}$ relation}
\label{sec:lxm}
Although $M_{\rm tot}$ depends on the adopted cosmology and, more strongly, on the group membership algorithm, $L_{\rm X}$ and $M_{\rm tot}$ remain among the least model-dependent tracers of AGN feedback. Figure \ref{flamingo_lxm} shows our covariance-corrected $L_{\rm X[0.1-2.4]}-M_{\rm tot}$ relation, with best-fit parameters listed in Table \ref{tab: relations}. We compared our results with BAHAMAS and FLAMINGO feedback models \citep{bras24}. Simulation luminosities, originally computed in the [0.5--2.0] keV band, were converted to our [0.1--2.4] keV band using a factor of 1.56 derived from an unabsorbed APEC model in \texttt{XSPEC}, assuming 0.4 Solar abundance, our median temperature (1.2 keV) and redshift (0.02), and the eROSITA response matrix. Following \citet{bras24}, we adopted \citet{asplund}'s solar abundances for consistency. The conversion factor ranges from 1.56 at 0.4 solar abundance to 1.35 for the extreme case of 1.9 Fe/H, but this difference is not significant ($<1\sigma$). We used core-excised $L_{\rm X}$ ($[0.15-1]R_{\rm 500c}$), which we estimated its contribution to be $\sim12\%$ for our sample. We accordingly removed this amount from the simulation luminosities.

As shown in the lower panel of Fig. \ref{flamingo_lxm}, BAHAMAS best reproduces the observed relation, agreeing within $1.0\sigma$ at our median mass, while the fiducial L1\_m9 FLAMINGO model is consistent at the $1.5\sigma$ level. In contrast, the stronger feedback models fgas-4$\sigma$ and fgas-8$\sigma$ underpredict $L_{\rm X}$ at the median mass of our sample, differing by $6.5\sigma$ and $8.6\sigma$, respectively. This agrees with \citet{eckert25}, who found similar tension with the observed $L_{\rm X}-M_{\rm tot}$ and $L_{\rm X}-T$ relations of the X-GAP sample \citep{xgap}. Figure \ref{lxm200} presents the covariance-corrected core-excised $L_{\rm X[0.1-2.4]}-M_{\rm tot}$ relation at $R_{\rm 200c}$, with best-fit parameters listed in Table \ref{tab: relations}.

\begin{figure}[hbt!]
    \centering
    \hspace{-0.1in}
    \includegraphics[height=0.28\textheight]{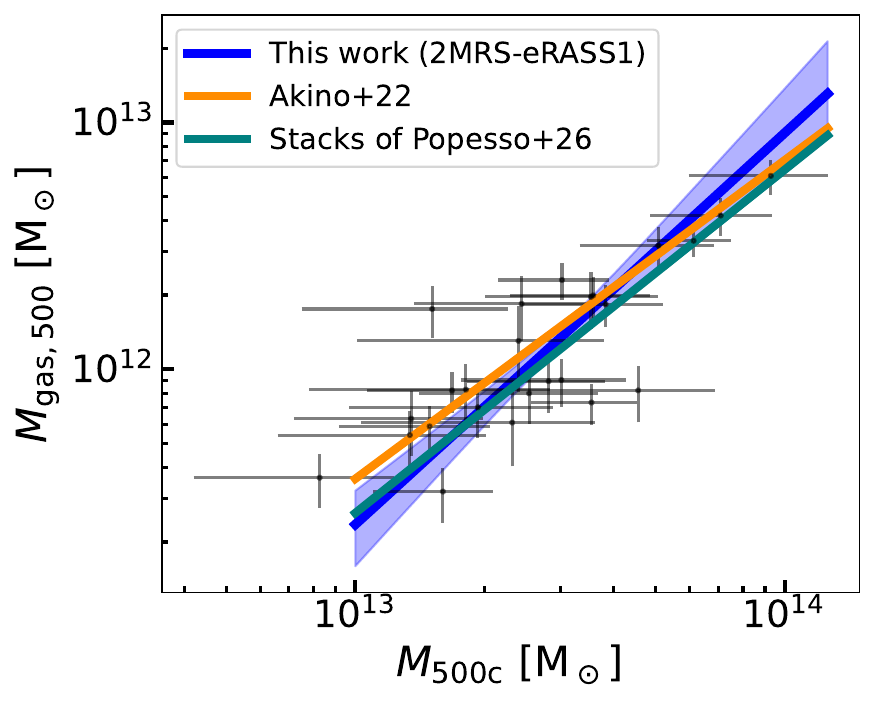}
    \caption{Gas mass vs total mass relation inside $R_{\rm 500c}$. The blue curve represents the best-fit relation for the 2MRS-eRASS1 sample. The blue shaded region is the $1\sigma$ confidence interval of the slope and normalisation. The orange curve indicates the result of the HSC-XXL sample \citep{akino22}, and the teal curve shows \citet{popesso_gas24}'s result from stacking eROSITA data of GAMA galaxy groups. The relation inside $R_{\rm 200c}$ is shown in Fig. \ref{mgmnfw200}.} 
    \label{mnfwmgas500}
\end{figure}

\begin{figure*}
    \centering
    \begin{subfigure}{0.9\textwidth}
        \centering
        \includegraphics[width=\linewidth]{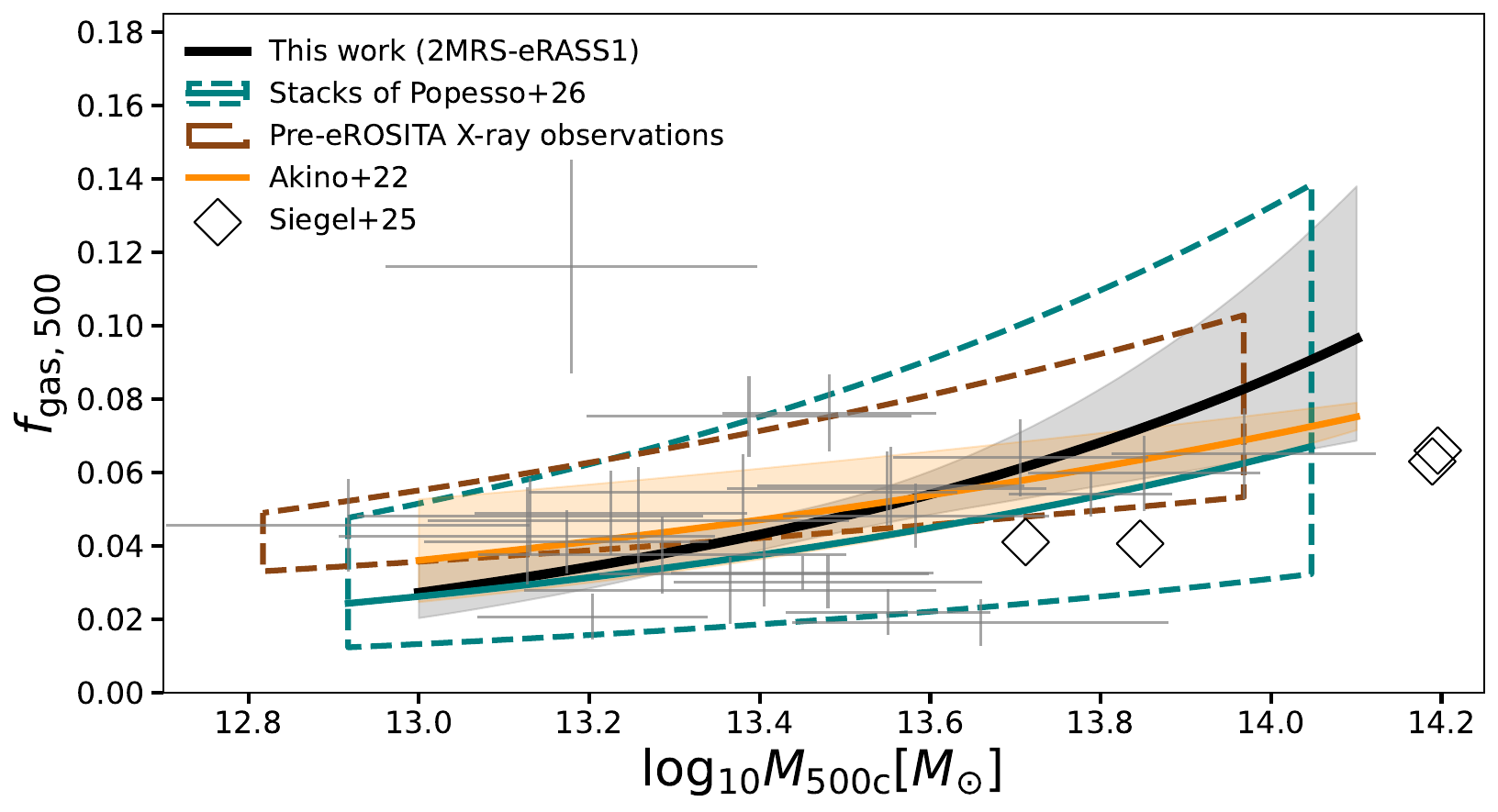}
    \end{subfigure}
    \vspace{0.1cm} 

    \begin{subfigure}{0.45\textwidth}
        \centering
        \includegraphics[width=\linewidth]{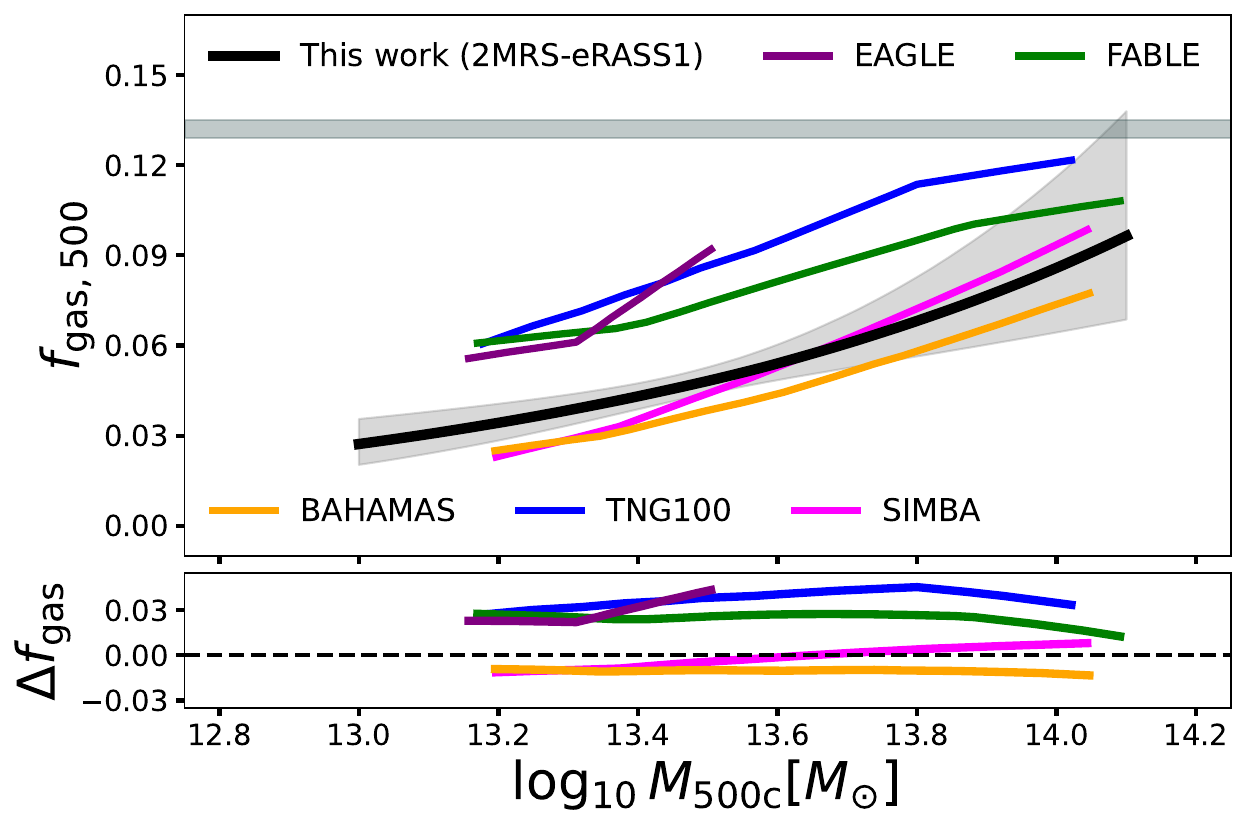}
    \end{subfigure}
    \hspace{0.01\textwidth}
    \begin{subfigure}{0.45\textwidth}
        \centering
        \includegraphics[width=\linewidth]{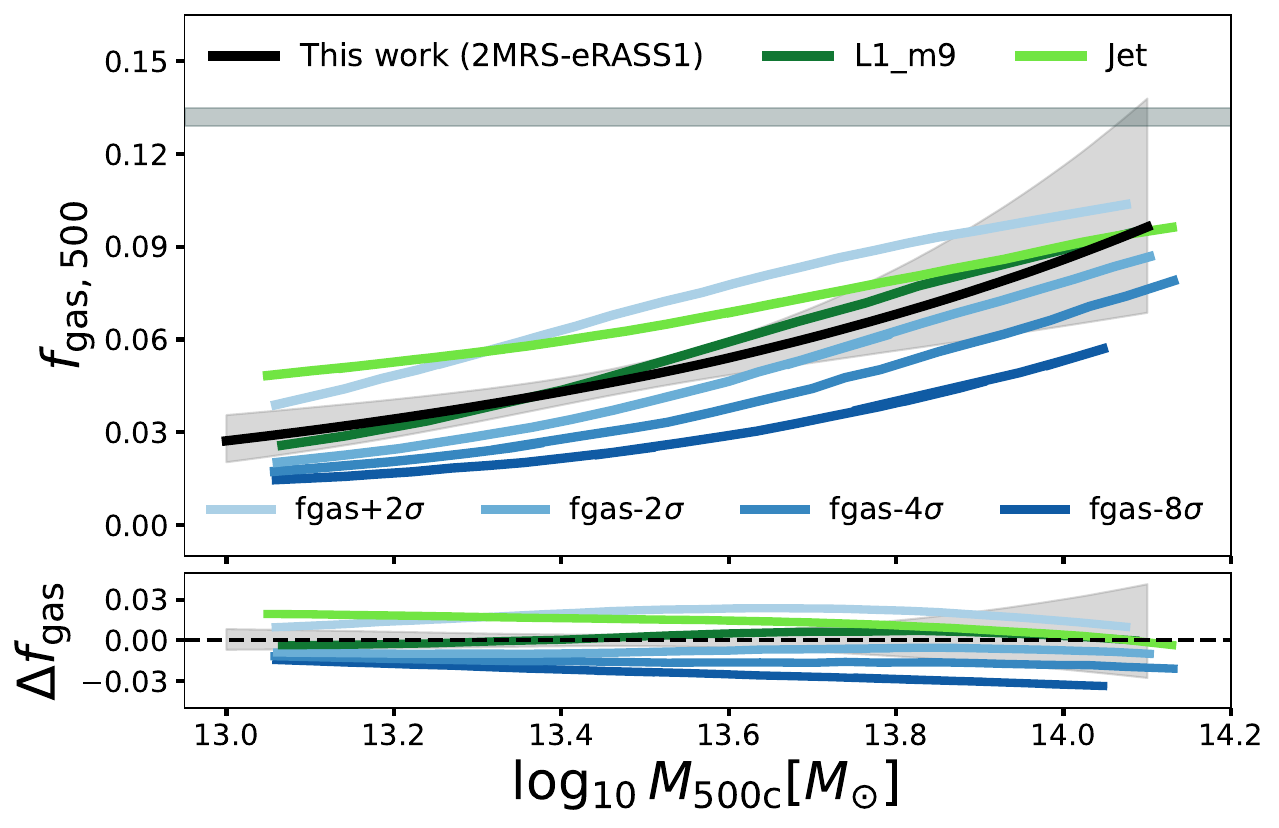}
    \end{subfigure}

     \caption{Hot gas fraction as a function of total mass inside $R_{\rm 500c}$. \textit{Top:} Comparison with observations. The black curve and shaded area show the  aperture covariance-corrected median relation and its $1\sigma$ uncertainty for our 2MRS-eRASS1 sample (Sect. \ref{sec: scaling relations}). The dashed brown area shows the pre-eROSITA X-ray compilation fit of \citet{eckert21}, the orange line and shaded area the HSC-XXL measurements \citep{akino22}, and the dashed~teal line and area the stacked eROSITA result for GAMA groups \citep{popesso_gas24}. Open diamonds show the kSZ+GGL+X-ray (DESI+ACT+eROSITA) stacking result of \citet{siegel25}. \textit{Bottom left:} Comparison with hydrodynamical simulations. Blue, purple, green, magenta, and orange show IllustrisTNG100 \citep{tng100}, EAGLE \citep{eagle}, FABLE \citep{fable18}, SIMBA \citep{simba}, and BAHAMAS \citep{bahamas17}, respectively. The dark grey slab indicates the cosmic gas fraction, and the lower subpanel shows residuals $\Delta f_{\rm gas}=f_{\rm sim}-f_{\rm 2MRS-eRASS1}$ vs $M_{\rm 500c}$. SIMBA best reproduces the observed relation. \textit{Bottom right:} Same as the bottom left but for FLAMINGO runs with various AGN feedback strengths; the fiducial L1\_m9 reproduces the observed $f_{\rm gas}-M_{\rm 500c}$ relation exactly.}
     \label{fgas500_m}
\end{figure*}

\subsubsection{$M_{\rm gas}-M_{\rm tot}$}
\label{mgas-m}
The gas mass versus\ total mass relation probes the hot gas budget in group atmospheres. In a self-similar scenario, the gas fraction reaches the cosmic value and a linear $M_{\rm gas}$--$M_{\rm tot}$ relation is expected, while a steeper slope signals non-gravitational processes expelling gas beyond the virial radius \citep{Lovisari15}. Our $M_{\rm gas, 500}$--$M_{\rm 500c}$ relation (Fig.~\ref{mnfwmgas500} and Table~\ref{tab: relations}) yields a median slope of $B = 1.59^{+0.34}_{-0.28}$, slightly steeper than self-similar, with intrinsic scatter $\sigma_{\mathrm{int}} = 0.23^{+0.12}_{-0.11}$. We compared our relation to \citet{akino22}, who studied 136 X-ray selected groups and clusters from the XXL survey \citep{xxl16} using weak-lensing masses from the Hyper Suprime-Cam (HSC) Subaru Strategic Program. We report consistency with the HSC-XXL sample within $1\sigma$ across the studied mass range. Comparison with \citet{popesso_gas24} shows perfect agreement in the low-to-intermediate mass regime, with deviations at the high-mass end.

We show the resulting covariance-corrected $f_{\rm gas, 500}-M_{\rm 500c}$ relation in Fig. \ref{fgas500_m}. To put our result in context, we compared it to a compilation of pre-eROSITA literature X-ray observations, as summarised by the fitting relation introduced by \citet{eckert21}, including the results on the gas fraction by \citet{xxl}, \citet{ettori15E}, \citet{pratt09}, \citet{chiu18}, \citet{Lovisari15}, \citet{sun09}, \citet{sanderson13}, \citet{nugent20}, and \citet{eckert19}. The relation reads
\begin{eqnarray}
   f_{\rm literature} = 0.079\pm0.03\times \left(\frac{M_{\rm 500c}}{10^{14}M_{\odot}}\right)^{0.22^{+0.06}_{-0.04}}.
\end{eqnarray}

We obtain an increasing $f_{\rm gas}$ with mass, in agreement with pre-eROSITA observations. We compared our results with recent stacking analyses by \citet{popesso_gas24} and \citet{siegel25}; the latter combines kinematic Sunyaev-Zel'dovich (kSZ) effect profiles from DESI/SDSS with eRASS1 X-ray data. We find a mild tension with \citet{siegel25}'s two independent stacked halo masses at the $1.4\sigma$ level at $\approx 10^{13.7}-10^{13.8}M_{\odot}$; their measured $f_{\rm gas,500}$ is lower than our 2MRS-eRASS1 results. Our relation is in excellent agreement with  \citet{akino22}.

\begin{figure}[hbt!]
\centering
   \includegraphics[width=0.47\textwidth]{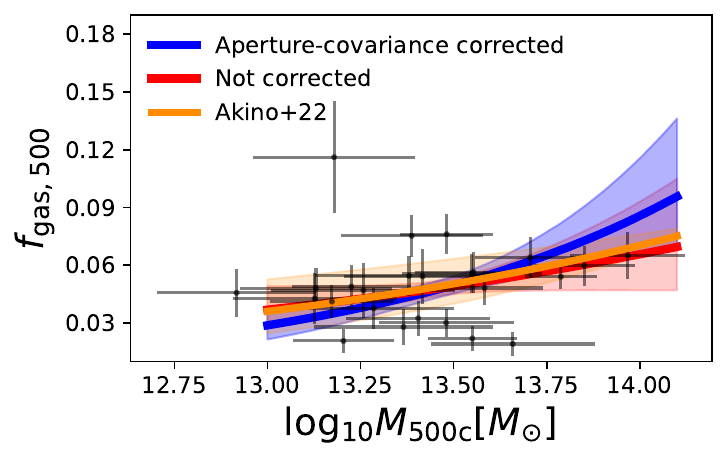}
     \caption{Aperture covariance correction. The corrected best-fit relation, shown also in Fig. \ref{fgas500_m}, is plotted in blue. We show the uncorrected relation with the red curve. The HSC-XXL result is shown in orange.}
     \label{covcorr}
\end{figure}

\citet{eckert21} compiled the predictions of the gas fraction at $R_{\rm 500c}$ by most modern hydrodynamical simulations as a function of $M_{\rm 500c}$, including BAHAMAS \citep{bahamas17}, EAGLE \citep{eagle}, FABLE \citep{fable18}, SIMBA \citep{simba}, and IllustrisTNG100 \citep{tng100}. We provide a comparison of our result to those predictions in the bottom left panel of Fig. \ref{fgas500_m}, where we add a residual plot showing the difference between our $f_{\rm gas}$ and that predicted by each simulation. We find that SIMBA and BAHAMAS best reproduce our 2MRS-eRASS1 gas fractions across the whole overlapping mass range with an average deviation of $0.64\sigma$ and $1.44\sigma$, respectively, with agreement improving at higher masses. SIMBA achieves $<1\sigma$ consistency at lower masses ($M_{\rm 500c}\approx10^{13.5}M_{\odot}$) than BAHAMAS ($M_{\rm 500c}\approx10^{13.8}M_{\odot}$), in agreement with the high-mass trend seen in \citet[see their Fig. 15]{eckert21}. At our median mass ($\log{M_{\rm 500c}}/M_{\odot}=13.4$), we find a $1.74\sigma$ deviation for SIMBA and $2.41\sigma$ for BAHAMAS. Conversely, IllustrisTNG100, FABLE, and EAGLE overpredict $f_{\rm gas,500}$ at $10^{13}<M_{\rm 500c} < 10^{14}M_{\odot}$, deviating from our relation by $7.3\sigma$, $7.1\sigma$, and $5.2\sigma$, respectively, at the median mass. 

Although SIMBA strongly underestimates the gas content in the inner regions relative to other simulations and is inconsistent with the observed inner hot intracluster medium properties (e.g. Fig. 2 in \citealt{xgap}), its $f_{\rm gas}$--$M$ relation converges with them at and beyond $R_{\rm 500c}$. Given the resolution limitations of the simulations considered, we focus our comparison on the larger scales of $R_{\rm 500c}$ and $R_{\rm 200c}$, and we find SIMBA to be consistent with our observed relation.

We also evaluate the predictions of FLAMINGO \citep{flamingo}, a simulation suite calibrated to observed nearby group gas fractions, in the bottom right panel of Fig. \ref{fgas500_m}. AGN feedback is implemented in FLAMINGO's model variants thermally by changing the temperature of gas particles around the black hole by $\Delta T_{\rm AGN}$, except in the Jet model, which uses kinetic acceleration of the particles ($v_{\rm jet}$). Both parameters are tuned to match cluster gas fractions \citep[see Table 4 in][]{kugel23}. The alternative fgas models (e.g. fgas$-8\sigma$, fgas$+2\sigma$) represent shifts in the calibrating data by corresponding multiples of their observational uncertainties.
We refer the reader to the reviews by \citet{eckert21} and \citet{oppen21} for details on AGN feedback implementation in simulations.

The fiducial FLAMINGO L1\_m9 model accurately reproduces our observed gas fractions between $10^{13} M_{\odot}$ and $10^{14} M_{\odot}$ with a negligible $0.17\sigma$ deviation at the median mass. However, strong feedback models significantly deviate from our relation with fgas$-8\sigma$, fgas$-4\sigma$, and fgas$-2\sigma$ models underestimating $f_{\rm gas}$ with tensions at our median mass of of $5.1\sigma$, $3.6\sigma$, and $2.2\sigma$, respectively. While the Jet and fgas$+2\sigma$ models overpredict the gas fraction by $3.9\sigma$ and $4.7\sigma$.

\begin{figure}[hbt!]
\centering
   \includegraphics[width=0.47\textwidth]{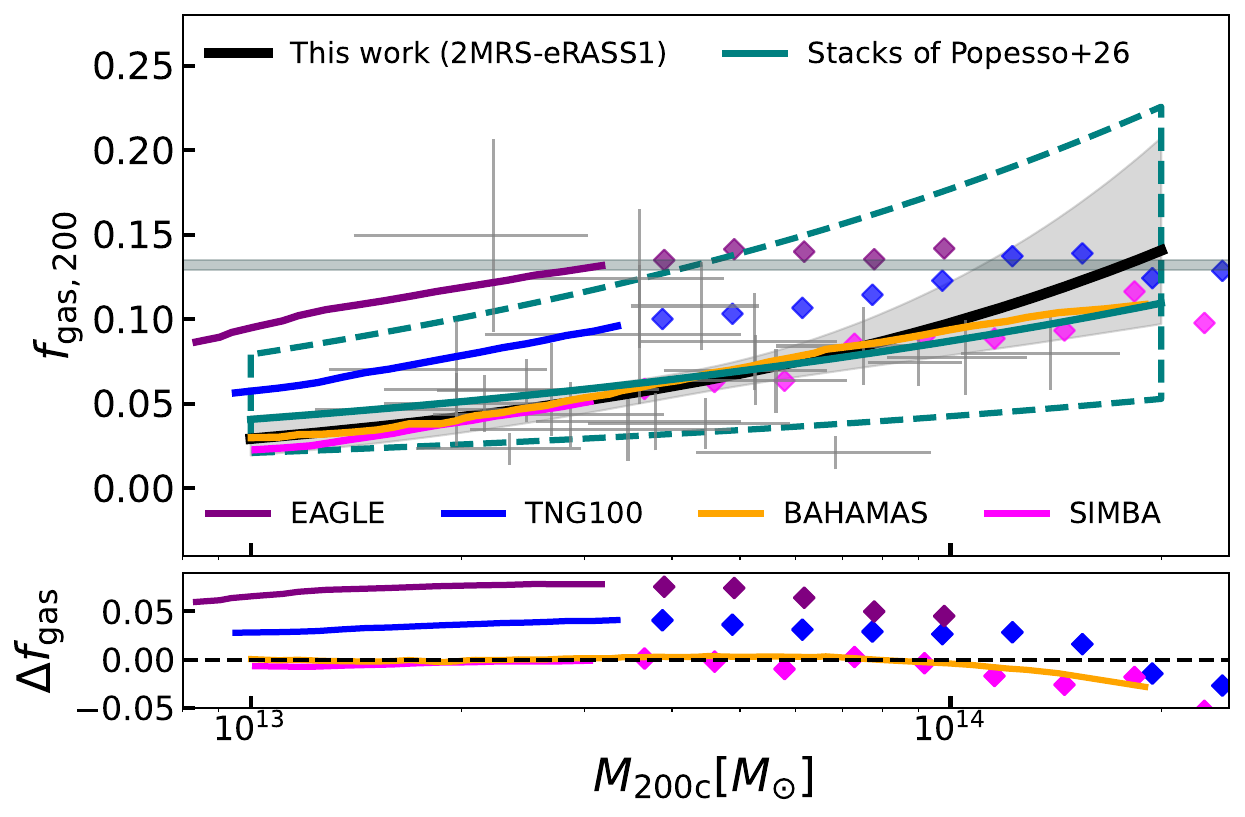}
     \caption{Hot gas fraction as a function of the total mass inside $R_{\rm 200c}$. The black curve represents the best-fit aperture covariance-corrected line to our 2MRS-eRASS1 sample. The purple, blue, orange, and magenta lines show the predictions of EAGLE \citep{eagle}, IllustrisTNG100 \citep{tng100}, BAHAMAS \citep{bahamas17}, and SIMBA \citep{simba} hydrodynamical simulations, respectively, as reported in \citet{oppen21}. Median simulation relations are shown up to a mass scale at which individual objects begin to be plotted. In this regime, we use 0.1 dex mass bins. The teal line indicates the eROSITA stacking result of \citet{popesso_gas24}. Other details are the same as in Fig. \ref{fgas500_m}.}
     \label{fgas200_m}
\end{figure}

In Sect. \ref{sec:aper} we formalise the aperture covariance effect for quantities with a shared integration radius, and provided a $\beta$-model slope-dependent correction. Figure \ref{covcorr} illustrates how correcting the negative $f_{\rm gas,500}-M_{\rm tot}$ correlation steepens the relation. We find that the flatter, uncorrected relation natively recovers the HSC-XXL results \citep{akino22}. 

Finally, we present the first individual-system eROSITA constraints on the $M_{\rm gas}-M_{\rm tot}$ relation at $R_{\rm 200c}$ (Fig. \ref{mnfwmgas500}, Table \ref{tab: relations}). The median slope ($B = 1.53$) is mildly shallower than at $R_{\rm 500c}$, indicating relatively more baryons at larger radii, which is expected from the density profile steepening beyond $R_{\rm 500c}$ (Fig. \ref{profiles}). Our $f_{\rm gas,200}-M_{\rm 200c}$ relation agrees with \citet{popesso_gas24} but with significantly higher precision (Fig. \ref{fgas200_m}). Comparing with simulations \citep{oppen21}, BAHAMAS and SIMBA excellently match our observations, whereas IllustrisTNG100 and EAGLE overpredict the gas fraction at the median mass by $\approx5\sigma$ and $10\sigma$, respectively.

\begin{figure*}[hbt!]
\centering
   \includegraphics[width=
   \textwidth]{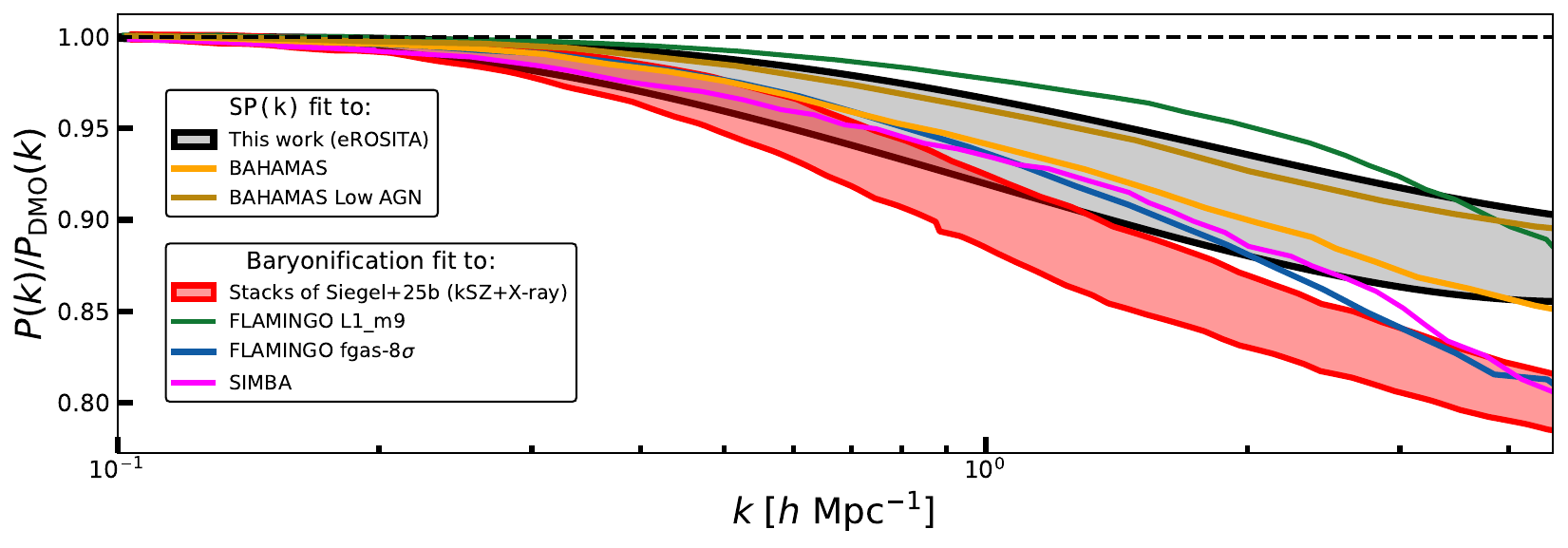}
     \caption{Suppression of the matter power spectrum. The black shaded area indicates our constraint on the power suppression implied by an \texttt{SP(k)} model \citep{salcido23} fit to the $f_{\rm b}-M$ relation of the 2MRS-eRASS1 sample. The orange and dark golden curves show \texttt{SP(k)} fits to the predictions of BAHAMAS and BAHAMAS Low AGN simulations, respectively. The red shaded region demonstrates a baryonification model \citep{schneider15, giri21} fit to stacked kSZ effect profiles combined with \textit{XMM-Newton} and eROSITA gas fractions \citep{siegel25b}. Baryonification model fits to the predictions of fiducial and strong-feedback FLAMINGO variants are shown with green and blue curves, respectively, while that to SIMBA is shown in magenta.}
     \label{mps}
\end{figure*}

\section{Matter power spectrum suppression}
\label{sec:mps}
The connection between the baryon fraction $f_{\rm b}$ in groups and clusters and the MPS suppression on large scales ($k < 10\ h\ \rm Mpc^{-1}$) is well established \citep[e.g.][]{schneider19,Debackere20,vandaalen2020,salcido23,schaller25}. We estimated the ratio of the MPS while considering baryonic physics to a dark matter only case $P(k)/P_{\rm DMO}(k)$ using the ANTILLES-based model \texttt{SP(k)} \citep{salcido23}, which predicts power suppression from the power-law-fitted parameters of the $f_{\rm b}-M$ relation at a given redshift and cosmology. ANTILLES comprises 400 simulations with sufficient volume and mass resolution to constrain baryonic effects on power suppression up to $k \lesssim 10\ h\ \rm Mpc^{-1}$ and $z\leq3$, incorporating a wide range of AGN feedback flavours spanning $-7\sigma$ to $+6\sigma$ relative to \citet{akino22}. \citet{salcido23} introduces the exponential plateau model \texttt{SP(k)} that approximates the MPS suppression from the baryon fraction measured at the optimal halo mass, defined as the mass at which the correlation between $f_{\rm b}$ and power suppression is maximised (see their Figs. 3 and 5).

To use \texttt{SP(k)}, we complemented our gas fractions with stellar mass fractions ($f_{\star}$)  from \citet{leauthaud12}, assuming a Chabrier initial mass function (IMF) and adopting $0.012 < f_{\star} < 0.025$ at our median $M_{\rm 500c}$. We re-fitted $(f_{\rm gas}+f_{\star})-M_{\rm 500c}$ with
    
       \begin{eqnarray}
        f_{\rm b,500}/(\Omega_{\rm b}/\Omega_{\rm m}) = \alpha \left(\frac{M_{\rm 500c}}{M_{\rm piv}}\right)^{\gamma}.
        \label{eq:mps}
    \end{eqnarray}
    
The systematic uncertainty of the stellar fraction\ $\sigma_{\star,\ \rm sys} \approx 0.25$ dex \citep{behroozi10, leauthaud12} is driven primarily by the IMF choice, with smaller contributions from dust attenuation, star formation history, and stellar population synthesis assumptions. In our analysis, we propagated $\sigma_{\star,\ \rm sys}$ alongside the gas fraction uncertainty when deriving the confidence interval. We stress that MPS suppression is sensitive to the total baryon budget and not just the gas fraction, as implied by the middle and right panels of Fig. 4 in \citet{siegel25b}. In that work, the SIMBA and fiducial FLAMINGO (L1\_m9) simulations yield similar gas fractions yet predict different suppression curves due to differing stellar, and thus total baryon, budgets (see Fig. 1 in \citealt{bigwood25}), highlighting the importance of accounting for all sources of error in stellar fraction estimates. 
    
Our constraints on $P(k)/P_{\rm DMO}(k)$ are shown in Fig. \ref{mps}, compared to \texttt{SP(k)} fits to fiducial and weak-AGN BAHAMAS models \citep{salcido23}, and to the baryonification model results of \citet{siegel25b} fitted to kSZ+\textit{XMM-Newton}+eROSITA data. On scales $k < 1\ h\ \rm Mpc^{-1}$, our constraint is in broad agreement with all hydrodynamical simulations and the kSZ+X-ray results. On smaller scales ($k \gtrsim 1.5\ h\ \rm Mpc^{-1}$), growing tension emerges with strong feedback FLAMINGO, SIMBA, and the kSZ+X-ray constraints, while fiducial FLAMINGO and BAHAMAS remain consistent throughout.

\section{Summary}
\label{sec:summary}
We have presented eROSITA X-ray properties of the hot intergalactic gas at the outskirts of a sample of 25 galaxy groups detected in the first public release of the eROSITA-DE data \citep[eRASS1;][]{merloni24, erass1} and confirmed in the 2MRS spectroscopic catalogue \citep{tempel18}. In our analysis, we used dynamical halo masses that were verified against the \citet{munari13} velocity dispersion-mass relation and gas masses obtained from a multi-scale de-projection of the EM profiles described in Sect. \ref{analysis: fgas}. We estimated the hot gas fraction profiles, up to $R_{\rm 200c}$, through a Monte Carlo process to propagate all sources of error. We constructed scaling relations between X-ray observables and the total mass and compared our results to stacking-based and individual-system observations, as well as to the predictions of most modern hydrodynamical simulations. Finally, we used our gas fraction results to measure the MPS suppression. We summarise our main findings as follows:

\begin{itemize}[label=\textbullet]
    \item 2MRS-eRASS1 groups exhibit uniformly flat SBx profiles that are well described by a $\beta$ model with a mean $\beta$ of $\sim 0.4$ and vanishing inner cores, such that the profiles rise as power laws towards the centre (Table \ref{table: beta}). This morphological property may  lead to the omission of such systems from cluster catalogues that rely on modelling X-ray peaks. Beyond $R_{\rm 500c}$, we detect a significant steepening, with $\beta_{[0.15-1]R_{\rm 500c}} = 0.38\pm0.04$ rising to $\beta_{R_{\rm 500c}-R_{\rm 200c}}= 0.76\pm0.19$ (Fig. \ref{profiles}).

   \item At the median mass $\log_{10}M_{\rm 500c}/M_{\odot}=13.4$, we measure $f_{\rm gas,500}=4.32\pm0.42\%$, which increases to $f_{\rm gas,200}=5.78\pm0.69\%$ at $R_{\rm 200c}$ and the median mass $\log_{10}M_{\rm 200c}/M_{\odot}=13.57$.\ However, it remains well below the cosmic value. This implies substantial AGN-driven gas ejection, consistent with \citet{ayromlou23}, who found mean closure radii of $2$, $3.5$, and $9.5\,R_{\rm 200c}$ in EAGLE, IllustrisTNG, and SIMBA, respectively. A detailed eROSITA study of the closure radius is the focus of a future study (Khalil et al. in prep.).

    \item We present the $M_{\rm gas}-M_{\rm tot}$, $L_{\rm X}-M_{\rm tot}$, and $L_{\rm X}-M_{\rm gas}$ scaling relations at $R_{\rm 500c}$ and $R_{\rm 200c}$ (Sect. \ref{sec: scaling relations}, Table \ref{tab: relations}). We corrected for aperture-induced covariance between quantities that share the same integration radius (Sect. \ref{sec:aper}) and modelled our optical selection.

    \item Our $f_{\rm gas}-M_{\rm tot}$ relation agrees with the stacking study of \citet{popesso_gas24} although with approximately twice the statistical precision at the median mass (Figs. \ref{fgas500_m} and \ref{fgas200_m}), is fully consistent with \citet{akino22} and \citet{eckert21}, and shows mild tension with \citet{siegel25}.
    
    \item Of the simulations (Figs.~\ref{flamingo_lxm}, \ref{fgas500_m}, and \ref{fgas200_m}), the fiducial FLAMINGO best reproduces both the $f_{\rm gas}$--$M_{\rm 500c}$ and $L_{\rm X}$--$M_{\rm 500c}$ relations. BAHAMAS and SIMBA provide the closest match to $f_{\rm gas,500}$--$M_{\rm 500c}$ over the full mass range, with FABLE and TNG100 agreeing only at the highest masses. At $R_{\rm 200c}$, EAGLE and TNG100 overpredict $f_{\rm gas,200}$ at the median mass. For the $L_{\rm X,500}$--$M_{\rm 500c}$ relation, strong FLAMINGO feedback models fgas$-4\sigma$ and fgas$-8\sigma$ deviate by $6.5\sigma$ and $8.6\sigma$ at the median mass, consistent with the independent findings of \citet{seppi26} based on FLAMINGO--X-GAP \citep{xgap} analogues.

    \item Combining our gas mass fractions with stellar mass fractions from \citet{leauthaud12} via the \texttt{SP(k)} model \citep{salcido23}, we infer a $10\%-15\%$ suppression of the MPS at $k = 5\ h\ \rm Mpc^{-1}$ relative to a dark-matter-only universe (Sect. \ref{sec:mps}, Fig. \ref{mps}). This is consistent with the fiducial FLAMINGO and BAHAMAS but in growing tension with strong feedback variants on smaller scales.
\end{itemize}

\begin{acknowledgements}
HK acknowledges support by the European Union’s Horizon Europe research and innovation programme under grant agreement No 101131928, project ACME. ET acknowledges funding from the HTM (grant TK202), ETAg (grant PRG3034) and the EU Horizon Europe (EXCOSM, grant No. 101159513).
 
\end{acknowledgements}

\bibliographystyle{aa} 
\bibliography{ref} 

\begin{appendix}

\section{Wavelet gallery}
The exposure-corrected wavelet images obtained according to the procedure outlined in Sect. \ref{sub:erass1} in the [0.6--2.3] keV band are shown in Fig. \ref{zoo_all}.
\label{wv_gallery}

\begin{figure*}[p]
    \centering
    \begin{subfigure}[b]{0.33\textwidth}
        \includegraphics[width=\textwidth, height=0.23\textheight]{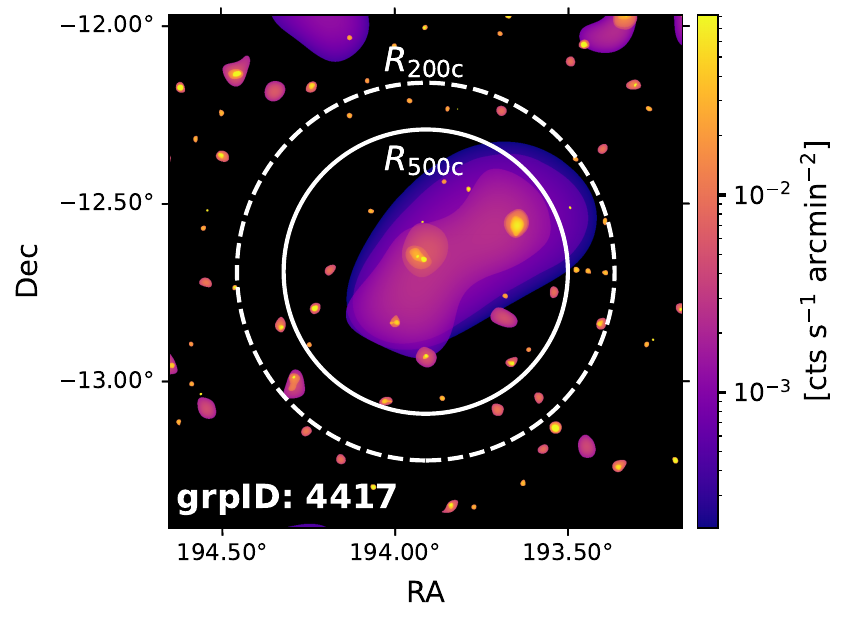}
    \end{subfigure}
    \begin{subfigure}[b]{0.33\textwidth}
        \includegraphics[width=\textwidth, height=0.23\textheight]{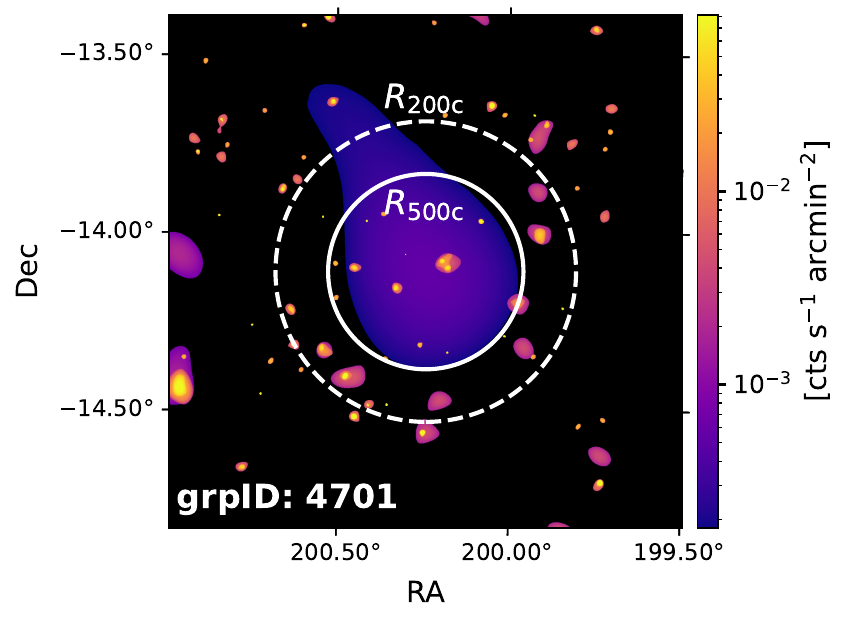}
    \end{subfigure}
    \begin{subfigure}[b]{0.33\textwidth}
        \includegraphics[width=\textwidth, height=0.23\textheight]{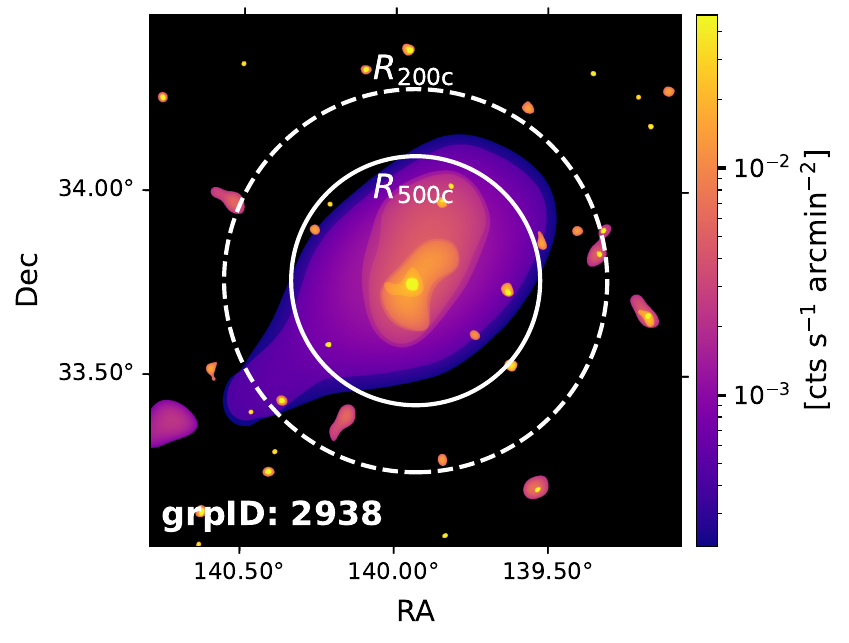}
    \end{subfigure}
    \begin{subfigure}[b]{0.33\textwidth}
        \includegraphics[width=\textwidth, height=0.23\textheight]{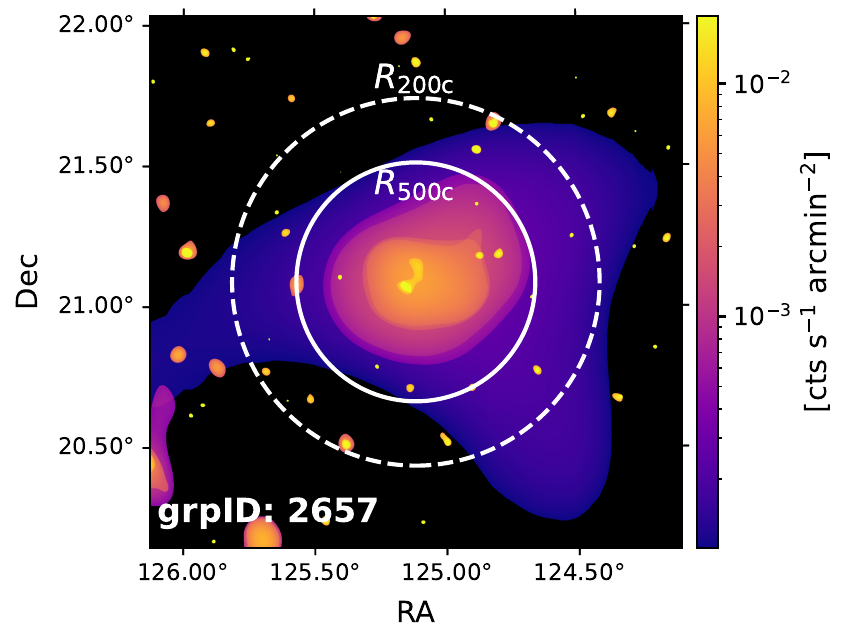}
    \end{subfigure}
    \begin{subfigure}[b]{0.33\textwidth}
        \includegraphics[width=\textwidth, height=0.23\textheight]{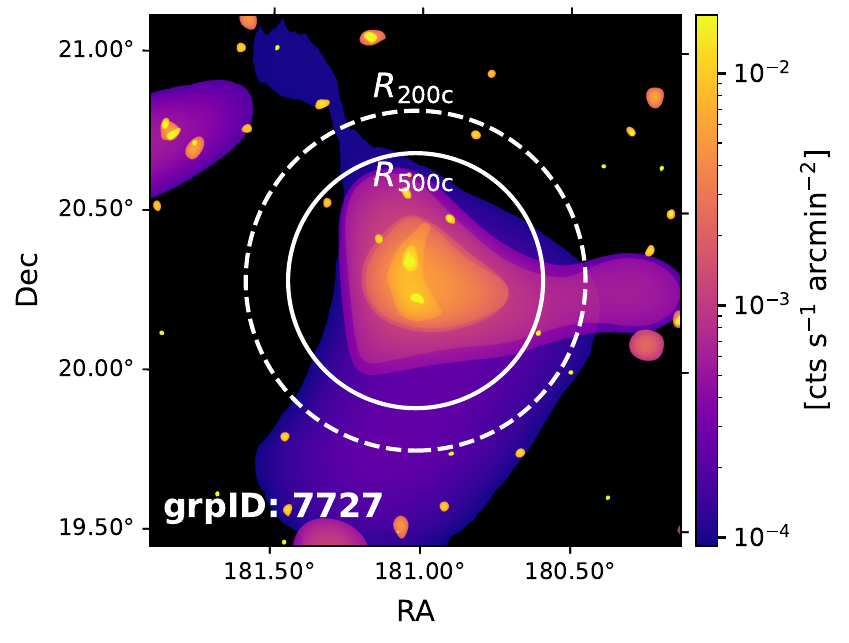}
    \end{subfigure}
    \begin{subfigure}[b]{0.33\textwidth}
        \includegraphics[width=\textwidth, height=0.23\textheight]{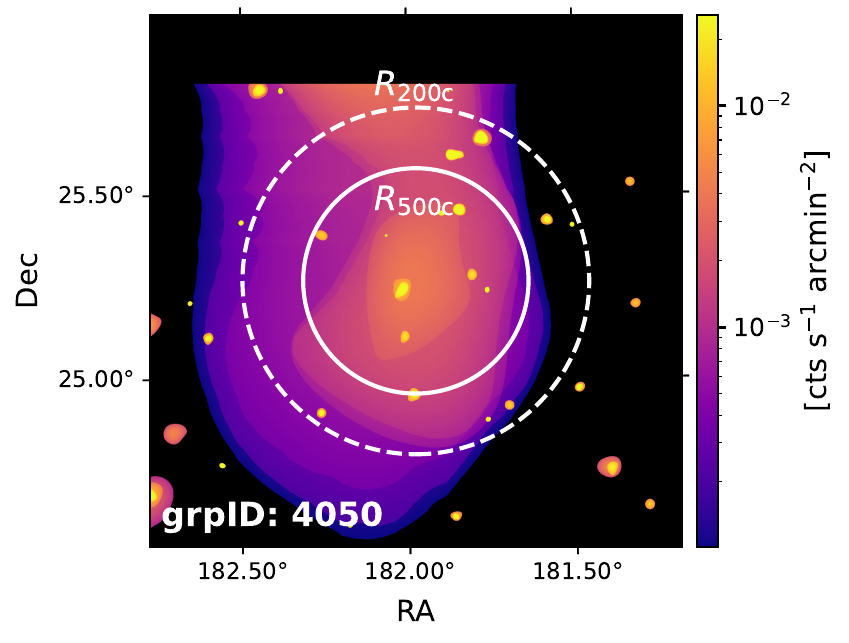}
    \end{subfigure}
    \begin{subfigure}[b]{0.33\textwidth}
        \includegraphics[width=\textwidth, height=0.23\textheight]{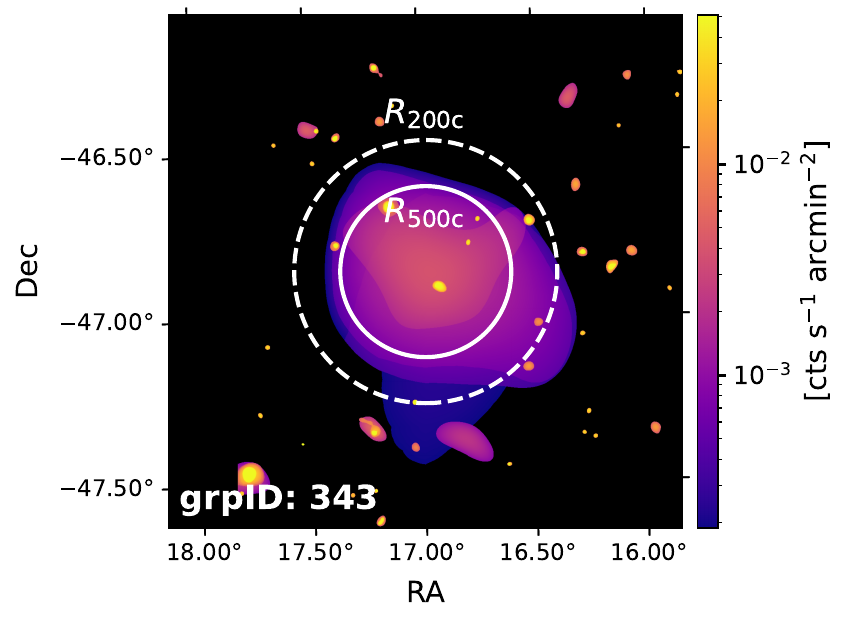}  
    \end{subfigure}
    \begin{subfigure}[b]{0.33\textwidth}
        \includegraphics[width=\textwidth, height=0.23\textheight]{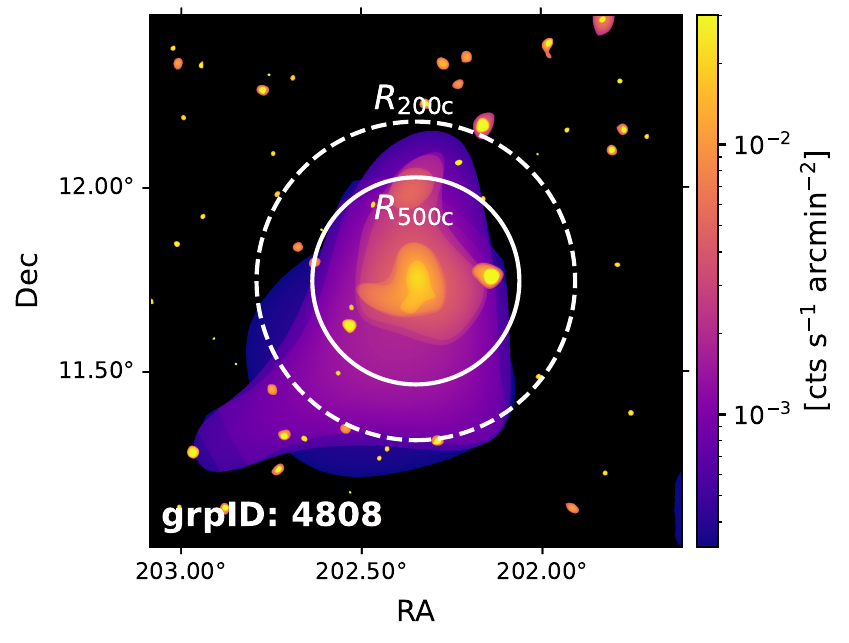}
    \end{subfigure}
    \begin{subfigure}[b]{0.33\textwidth}
        \includegraphics[width=\textwidth, height=0.23\textheight]{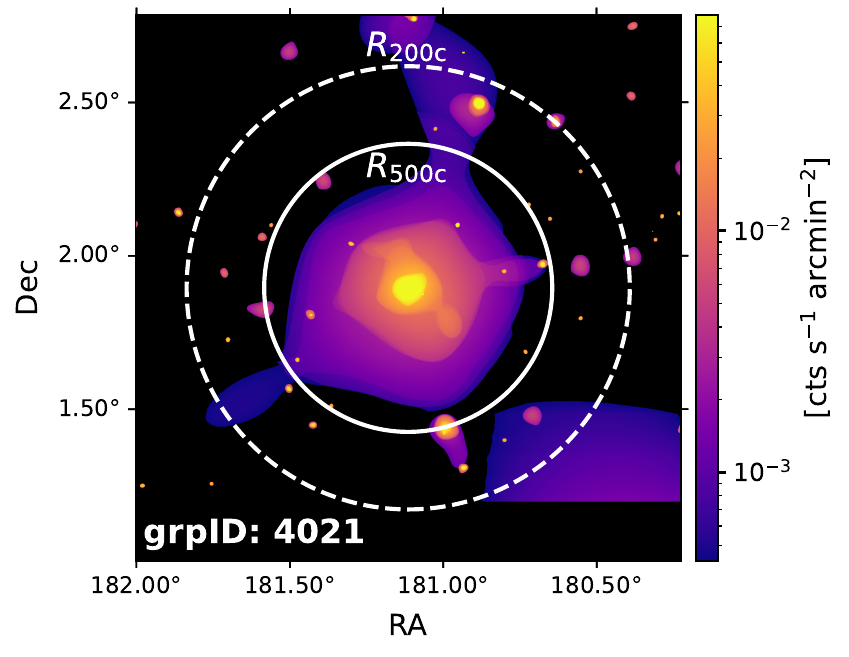}
    \end{subfigure}
    \begin{subfigure}[b]{0.33\textwidth}
        \includegraphics[width=\textwidth, height=0.23\textheight]{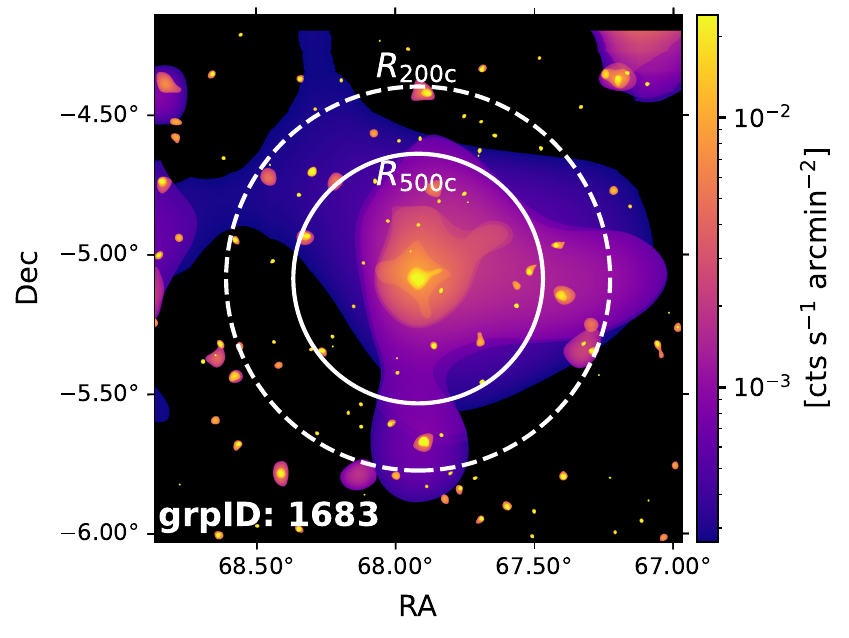}
    \end{subfigure}
    \begin{subfigure}[b]{0.33\textwidth}
        \includegraphics[width=\textwidth, height=0.23\textheight]{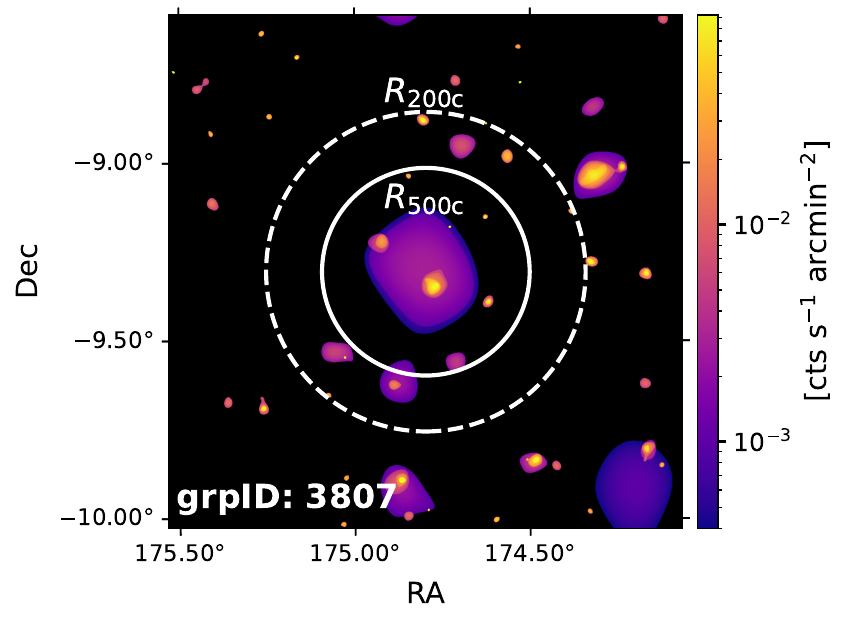}
    \end{subfigure}
    \begin{subfigure}[b]{0.33\textwidth}
        \includegraphics[width=\textwidth, height=0.23\textheight]{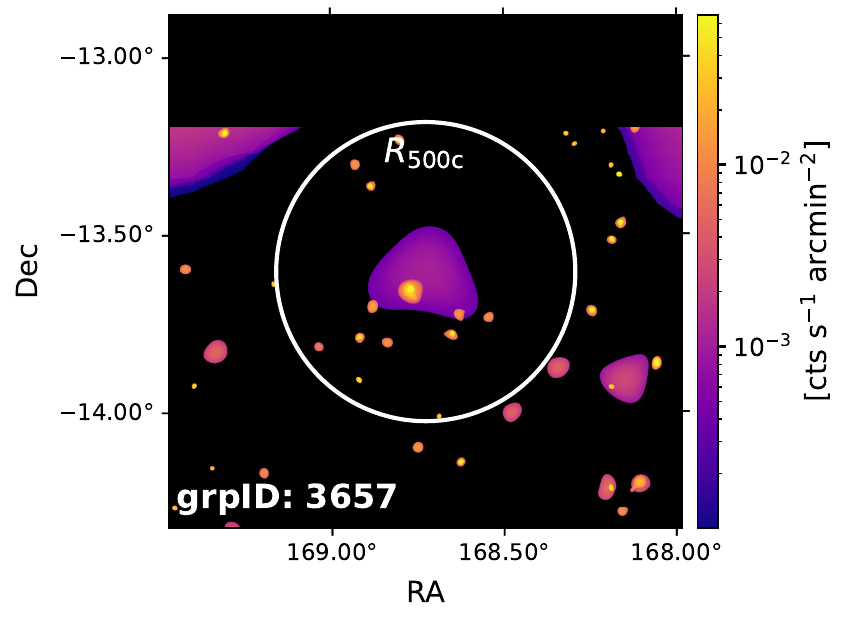}
    \end{subfigure} 
    \caption{Exposure-corrected [0.6-2.3] keV band eROSITA TM0 wavelet maps of the 2MRS-eRASS1 group sample. The 2MRS group identifier is indicated in each cutout, and the radial extent of $R_{\rm 500c}$ and $R_{\rm 200c}$ is shown with solid and dashed circles, respectively. The colour bar is reported in units of $\rm{cts}/s/arcmin^{2}$.}
    \label{zoo1}
\end{figure*}

\begin{figure*}[hbt!]
    \ContinuedFloat
    \centering
    \begin{subfigure}[b]{0.33\textwidth}
        \includegraphics[width=\textwidth, height=0.23\textheight]{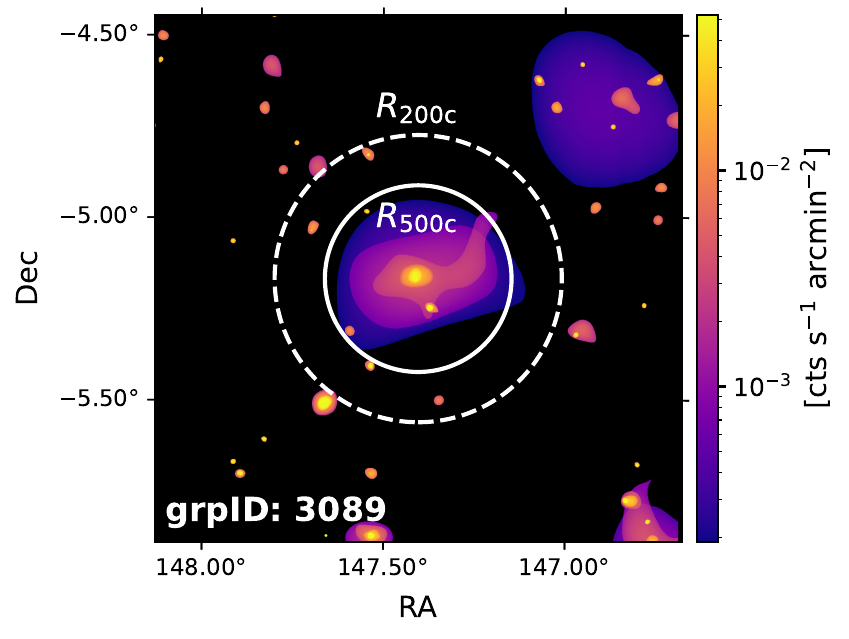}
    \end{subfigure}
    \begin{subfigure}[b]{0.33\textwidth}
        \includegraphics[width=\textwidth, height=0.23\textheight]{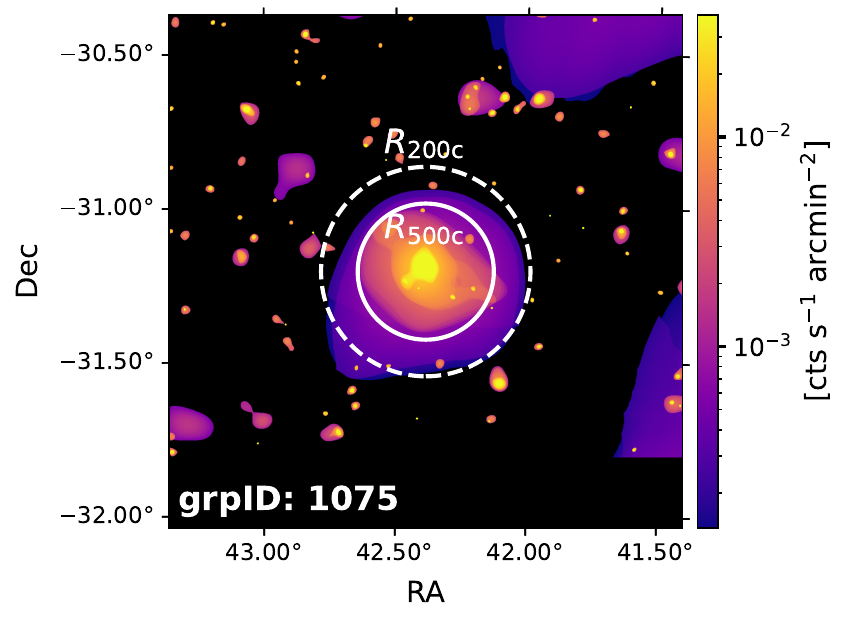}
    \end{subfigure}
    \begin{subfigure}[b]{0.33\textwidth}
        \includegraphics[width=\textwidth, height=0.23\textheight]{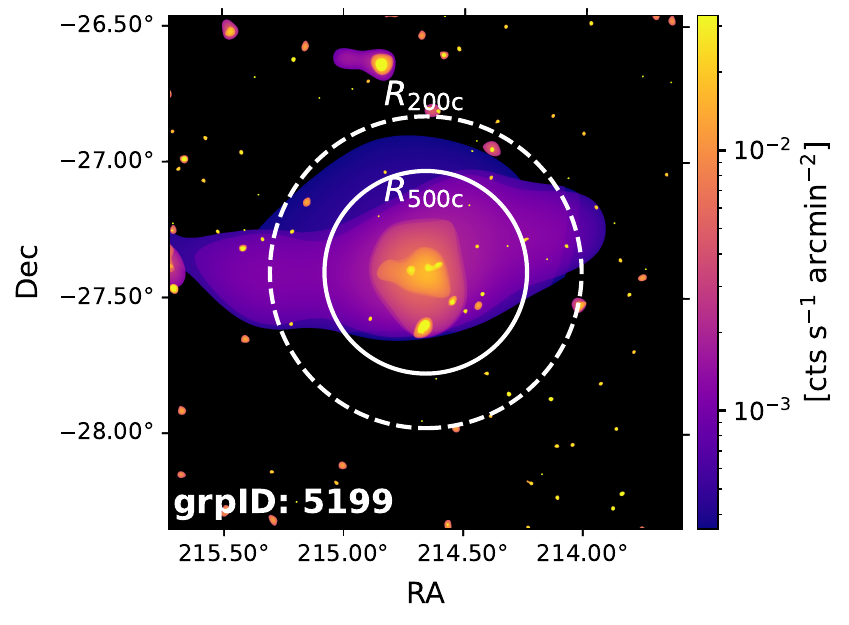}
    \end{subfigure}
    \begin{subfigure}[b]{0.33\textwidth}
        \includegraphics[width=\textwidth, height=0.23\textheight]{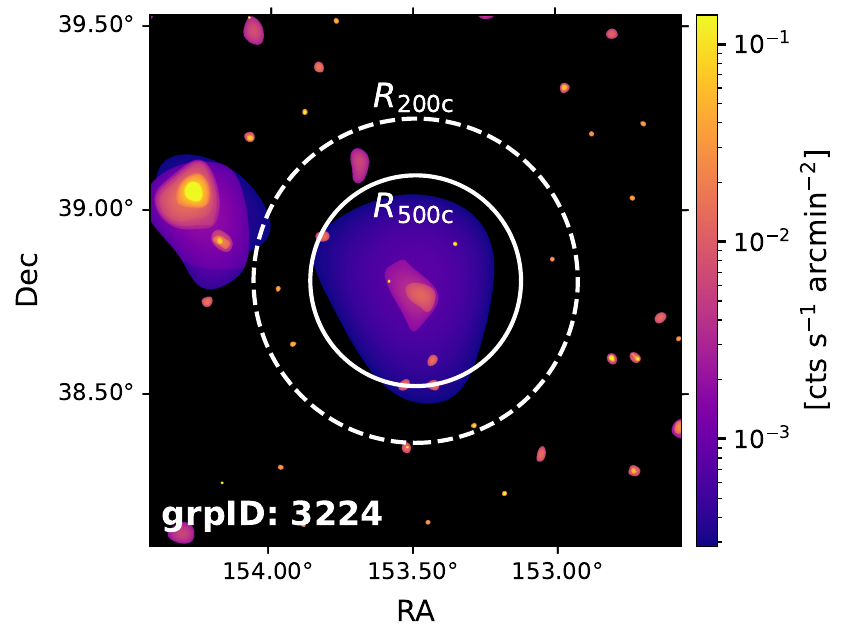}
    \end{subfigure}
    \begin{subfigure}[b]{0.33\textwidth}
        \includegraphics[width=\textwidth, height=0.23\textheight]{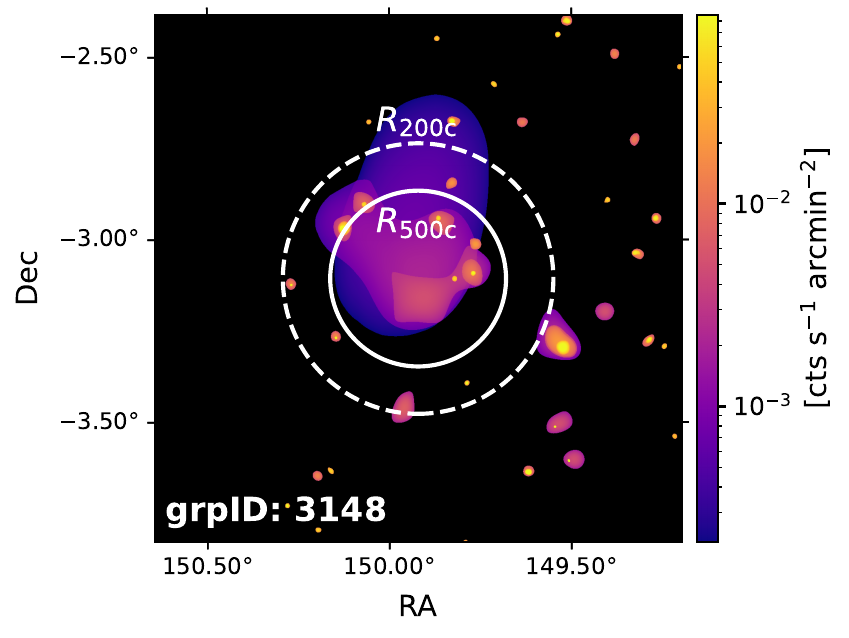}
    \end{subfigure}
    \begin{subfigure}[b]{0.33\textwidth}
        \includegraphics[width=\textwidth, height=0.23\textheight]{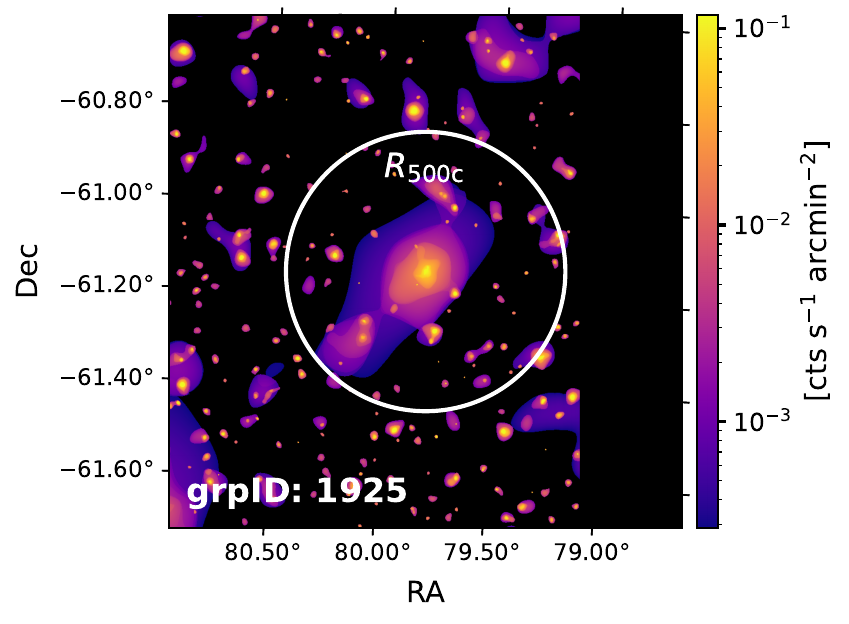}
    \end{subfigure}
    \begin{subfigure}[b]{0.33\textwidth}
        \includegraphics[width=\textwidth, height=0.23\textheight]{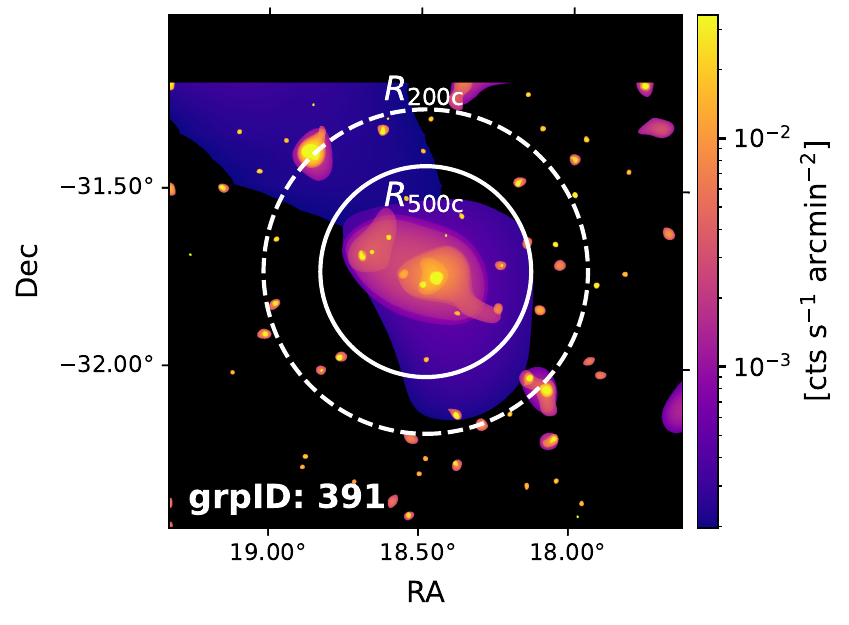}    
    \end{subfigure}
    \begin{subfigure}[b]{0.33\textwidth}
        \includegraphics[width=\textwidth, height=0.23\textheight]{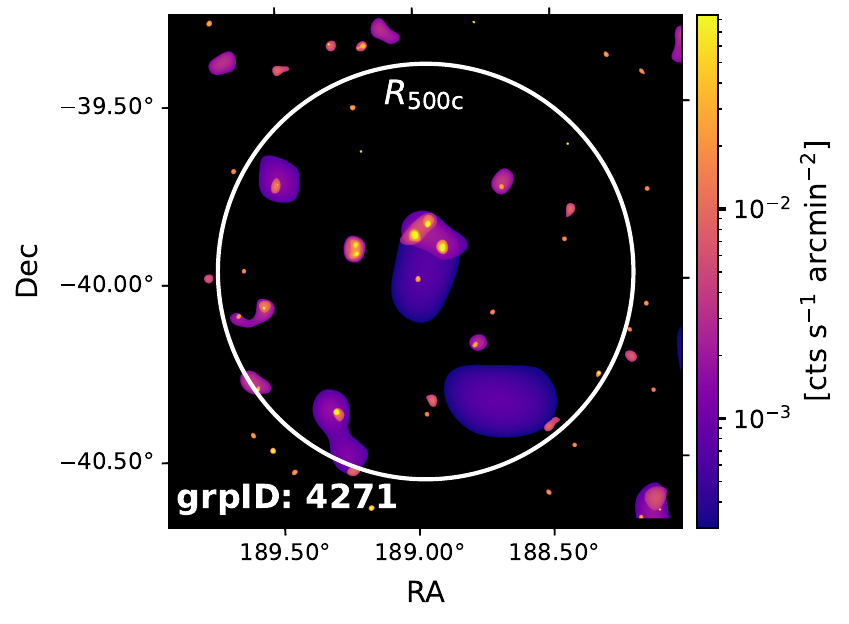}
    \end{subfigure}
    \begin{subfigure}[b]{0.33\textwidth}
        \includegraphics[width=\textwidth, height=0.23\textheight]{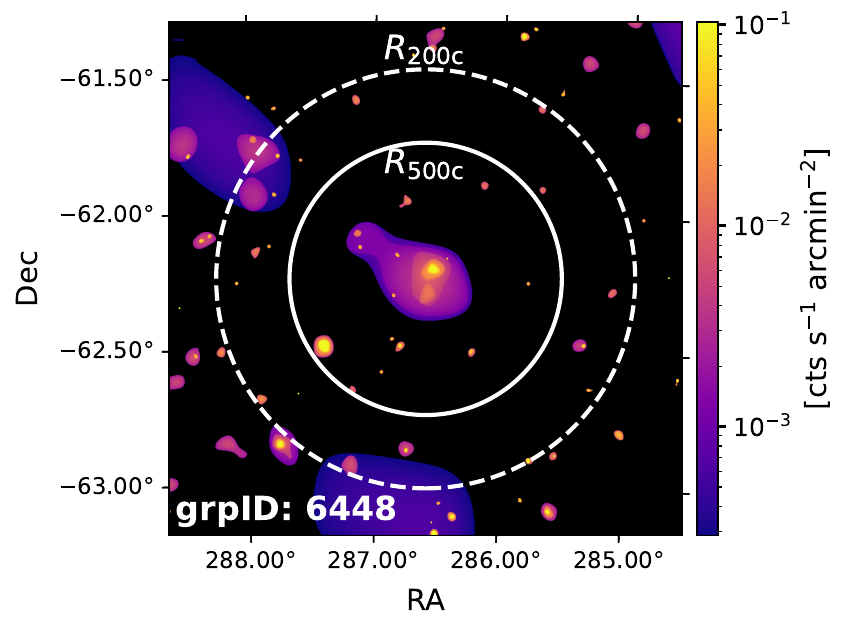}
    \end{subfigure}
    \begin{subfigure}[b]{0.33\textwidth}
        \includegraphics[width=\textwidth, height=0.23\textheight]{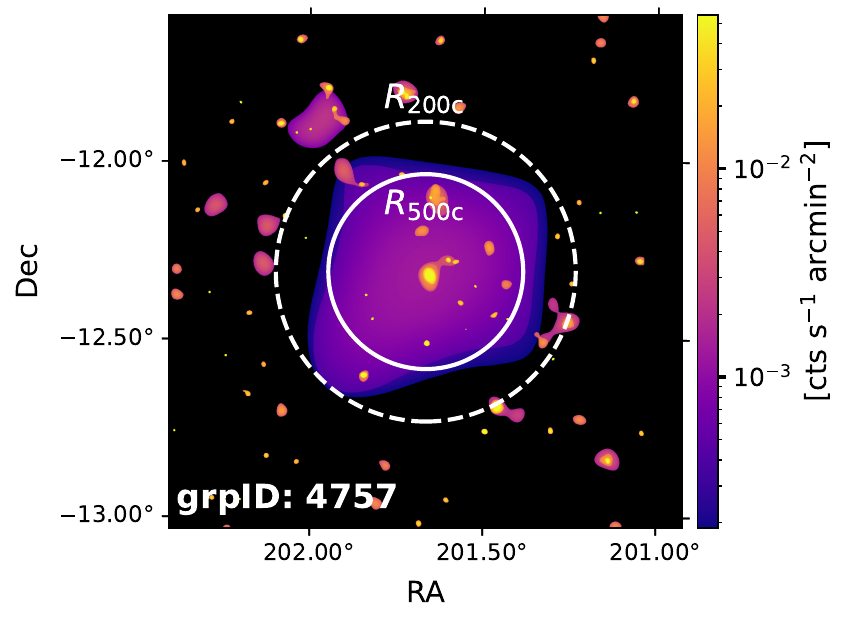}
    \end{subfigure}
    \begin{subfigure}[b]{0.33\textwidth}
        \includegraphics[width=\textwidth, height=0.23\textheight]{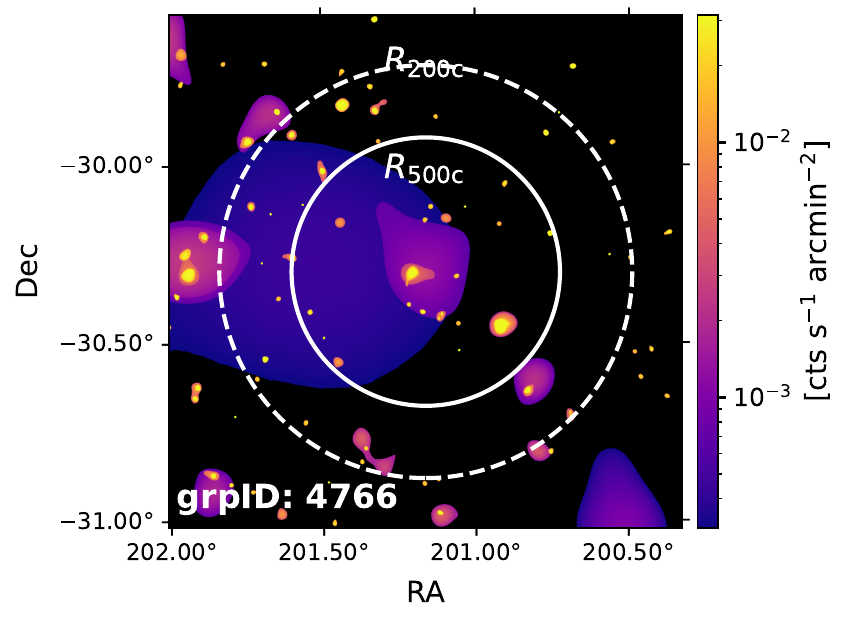}
    \end{subfigure}
    \begin{subfigure}[b]{0.33\textwidth}
        \includegraphics[width=\textwidth, height=0.23\textheight]{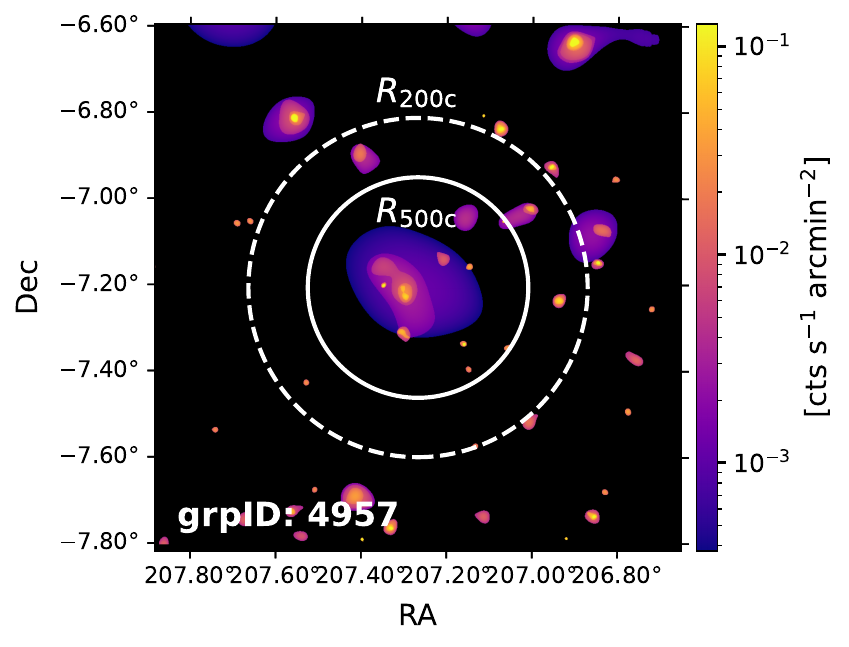}
    \end{subfigure}
    \caption{Continued.}
    \label{zoo2}
\end{figure*}

\begin{figure}[hbt!]
    \ContinuedFloat
    \centering
    \begin{subfigure}[b]{0.33\textwidth}
            \includegraphics[width=\textwidth, height=0.23\textheight]{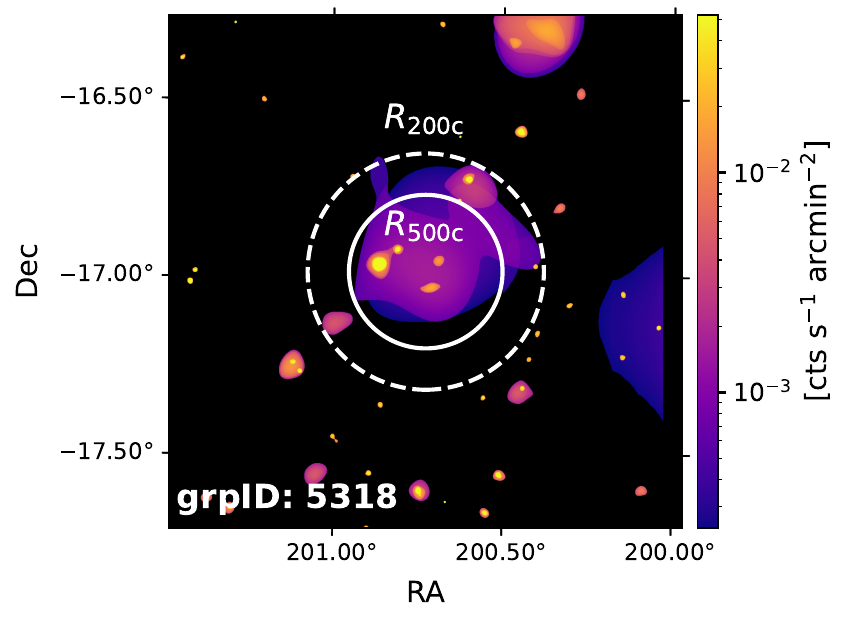}
        \end{subfigure}
    
    \caption{Continued.}
    \label{zoo_all}
\end{figure}

\section{Particle-induced background}
\label{section:pib}
Particle-induced background maps were generated for each sky tile following the procedure outlined in \citet{pib}. We summarise the procedure as follows: due to the stability of the PIB spectral shape and its negligible spatial variation, we measure the count rate in a hard energy band [$6-9$] keV, which is dominated by PIB events, and scale it to the [$0.6-2.3$] keV band of interest using a fixed ratio $R = S_{\rm PIB[0.6-2.3]keV}/H_{\rm PIB[6-9]keV}$ derived from the FWC\footnote{\url{https://eROSITA.mpe.mpg.de/dr1/AllSkySurveyData_dr1/FWC_dr1/}} data. The final PIB map is created by multiplying the hard band counts of the observation $H_{\rm obs}$ by $R$ and spatially distributing the product using the unvignetted exposure maps created for the [$0.6 - 2.3$] keV band using the eSASS task, \texttt{expmap}. The values of $S_{\rm PIB[0.6-2.3] keV}$ and $H_{\rm [6-9]keV}$ of the FWC data are listed in Table \ref{table: fwc}.

\begin{table}[hbt!]
\caption{Number of soft band [$0.6-2.3$] keV and hard band [$6-9$] keV counts in eROSITA FWC data.}
\label{table: fwc}
\centering
\begin{tabular}{c c c}
\hline\hline
TM & $S_{\rm PIB[0.6-2.3]}$ & $H_{\rm PIB[6-9]}$\\
\hline
1 & 62011 & 99243\\
2 & 44676 & 68931\\
3 & 46737 & 72632\\
4 & 46425 & 77307\\
5 & 25518 & 40623\\
6 & 47562 & 77529\\
7 & 50560 & 82620\\
\hline
\end{tabular}
\end{table}

\section{eROSITA point spread function}
\label{sec:psf}
Modelling the shape of the eROSITA survey PSF is required to take into account the effect of photon smearing across the inner annuli of the galaxy group emission and accurately characterise core radii of the emission region. We performed an in-flight calibration by searching eRASS1 on point sources with high detection likelihood ($\texttt{DET\_LIKE\_0}$) and 0 extent likelihood ($\texttt{EXT\_LIKE}$). Next, radial SBx profiles are extracted for three such point sources using logarithmically spaced concentric annuli with a minimum bin size of 1.5 arcsec and extending out to 4 arcmin. The coordinates of those point sources, their eROSITA sky tiles, and source identifiers within each tile, are listed in Table \ref{table: point_sources}. We then fitted them with a King model of the form
\begin{eqnarray}
   \Sigma(r) = \Sigma_{0} \left[ 1+ \left( \frac{r}{R_{\rm c}} \right)^{2} \right]^{-\alpha},
   \label{eq:beta}
\end{eqnarray}
where $\Sigma_{0}$, $r_{\rm c}$, and $\alpha$ are the normalisation, the core radius, and the slope of the King model, respectively. Our eROSITA PSF model has mean best-fit parameters of $R_{\rm c} = 6$ arcsec and $\alpha = 1.19$. To verify our PSF model, we integrated Eq. \ref{eq:beta} with these best-fit parameters out to a large radius of $100$ arcmin to estimate the total flux. We then repeated this integration out to the eROSITA survey half-energy width of $30$ arcsec in the [0.2--2.3] keV energy band \citep{merloni24}, and retrieved exactly $50\%$ of the total flux, confirming the validity of our PSF model. We assessed the effect of stray light from single mirror reflections on our analysis by cross-checking our results using the PSF model introduced in Appendix A in \citet{churazov23}. The difference in the observed SBx between this model and our simple one is $\sim1\%$ in regions of interest $\sim20^{\prime}$, and thus we found no impact on the gas mass results. 

\begin{table}[hbt!]
\caption{AGNs used in the PSF calibration.}
\label{table: point_sources}
\centering
\begin{tabular}{c c c c}
\hline\hline
eROSITA tile & Src ID & RA (deg) & Dec (deg)\\
\hline
085153 & 1 & 84.73 & -64.09\\
202069 & 2 & 201.22 & 21.60\\
006135 & 1 & 6.25 & -45.50\\
\hline
\end{tabular}
\end{table}

\section{Background regions}
\label{sec:bkg}

The X-ray background in each eROSITA tile was estimated from nine circular regions, each with 15-arcmin radius, sampling a range of FOV locations and distances from the source, to mitigate spatial background variability (example in Fig.~\ref{zoomed} and the mean background SBx per field listed in Table~\ref{table: beta}). We utilised knowledge of the extended wavelet scales to avoid placing background apertures in regions of extended X-ray emission (such as those associated with infalling substructure).

\begin{figure}[hbt!]
\centering
   \includegraphics[width=0.5\textwidth]{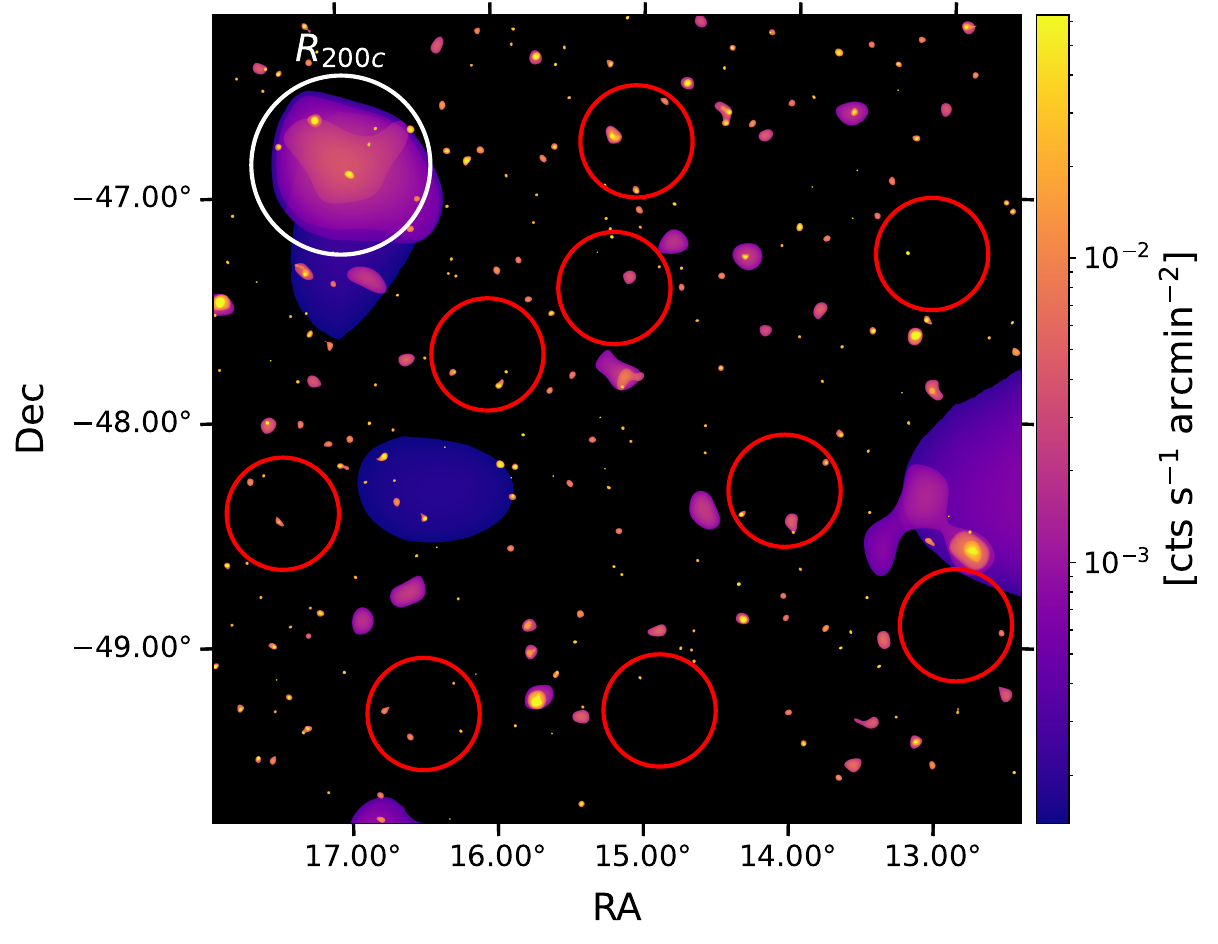}
     \caption{Source and background extraction regions of a 2MRS-eRASS1 example galaxy group. The white circle indicates the extent of $R_{\rm 200c}$. For each group, we define nine 15$'$ background extraction regions (red circles), selected manually to explore the variation in the background across the FOV, and we adopt the median value.}
     \label{zoomed}
\end{figure}

\section{Literature comparison}
\label{sec:comp}

To validate our analysis, we compared our derived gas masses with literature results for the same groups using different X-ray instruments (\textit{Chandra} and \textit{XMM-Newton}) at the same physical radius. We found literature studies for seven of our 25 groups, which we detail below, and show the comparison in Fig. \ref{comp_lit}. Overall, we report consistent gas mass measurements with previous instruments with a systematic $15\pm4.5\%$ lower gas mass values using eROSITA, a trend also seen for $L_{\rm X}$ in Fig. 18 in \citet{erass1}.

\begin{itemize}
    \item 2MRS 2657 (NGC 2832): \citet{gastaldello07} measured a gas mass of $3.1\pm0.3 \times 10^{11}M_{\odot}$ at 185 kpc with \textit{XMM-Newton} European Photon Imaging Camera (EPIC pn) versus our $3.5\pm0.7\times10^{11}M_{\odot}$. At 456 kpc, \citet{morandi17} used \textit{Chandra} Advanced CCD Imaging Spectrometer (ACIS-I)  to measure a gas mass of $2.89\pm0.2\times10^{12}M_{\odot}$ (vs our $2.19\pm0.4\times10^{12}M_{\odot}$) and $4.5 \pm 1.0 \times10^{12} M_{\odot}$ at 691 kpc (vs our $2.92 \pm 0.53 \times 10^{12} M_{\odot}$).
    \item 2MRS 2938 (NGC 2832): \citet{goulding16} reported the Chandra ACIS-I X-ray properties of only the central galaxy in a $21^{\prime\prime}$ aperture with no data found on larger scales.
    \item 2MRS 7727 (IC 4320): \citet{sutton15} reported on a hyper-luminous X-ray source candidate (2XMM J120405.8+201345) associated with the galaxy IC 4320 using \textit{Chandra} ACIS-S with no gas measurements on larger scales.
    \item 2MRS 4808 (NGC 5171): \citet{osmond04} measured the \textit{XMM-Newton} $L_{\mathrm{ X,500}}$ in the [0.4--2.0] keV band to be $3.47\pm0.7\times10^{42}$ erg s$^{-1}$, which corresponds to $5.76\pm 1.1\times10^{42}$ erg s$^{-1}$ in our [0.1--2.4] keV band of interest (consistent with our $5.01\pm0.39\times10^{42}$ erg s$^{-1}$ value).
    \item 2MRS 1683 (NGC 1600): This group meets many of the fossil group criteria such as having a high dark matter concentration \citep{runge22} and a large magnitude gap between it and its nearest neighbours \citep{smith08}. However, it lacked the high $L_{\rm X}$ associated with fossil groups ($L_{\rm X} > 5\times 10^{41}$ erg s$^{-1}$) as its previously measured $L_{\rm X}$  with \textit{Chandra} is $2.9\times10^{41}$ erg s$^{-1}$ \citep{sivakoff04}. We present the first eROSITA observation of this group with core-excised $L_{\rm X,500} = 2.46\pm0.2\times10^{42}$ erg s$^{-1}$, almost an order of magnitude higher than previously reported, resolving its fossil group $L_{\rm X}$ discrepancy. 
    \item 2MRS 1075 (IC 1860): This system is the only group with super-cosmic $f_{\rm gas,200}$ and is an outlier in the $f_{\rm gas}-M_{\rm 500c}$ relation. \citet{gastaldello13} show that IC 1860 has two SBx discontinuities (cold fronts) due to a minor merger, which explains its unusually high $f_{\rm gas}$. At $319$ kpc,\citet{gastaldello07} reported a gas mass of $1.50\pm0.05\times10^{12}M_{\odot}$ with \textit{XMM-Newton} (vs our $1.56\pm0.32\times10^{12}M_{\odot}$).
    \item 2MRS 4021 (MKW4): At $353$ kpc, \citet{gastaldello07} reported a gas mass of $2.84\pm0.06\times10^{12}M_{\odot}$ with \textit{Chandra} ACIS-S (vs our $2.20\pm0.35\times10^{12}M_{\odot}$). At $538$ kpc, \citet{sun09} used \textit{Chandra} ACIS to measure a gas mass of $4.17\pm0.38\times10^{12}M_{\odot}$ (vs our $4.3\pm 0.69\times10^{12}M_{\odot}$. At 884 kpc, \citet{sarkar21} measured a value of $8.73\pm0.7\times10^{12}M_{\odot}$ using a combined \textit{Chandra} ACIS-I + \textit{Suzaku} XIS analysis (vs our $8.62\pm1.38\times10^{12}M_{\odot}$).  
 
\end{itemize}

\begin{figure}[hbtp!]
\centering
\includegraphics[width=0.4\textwidth]{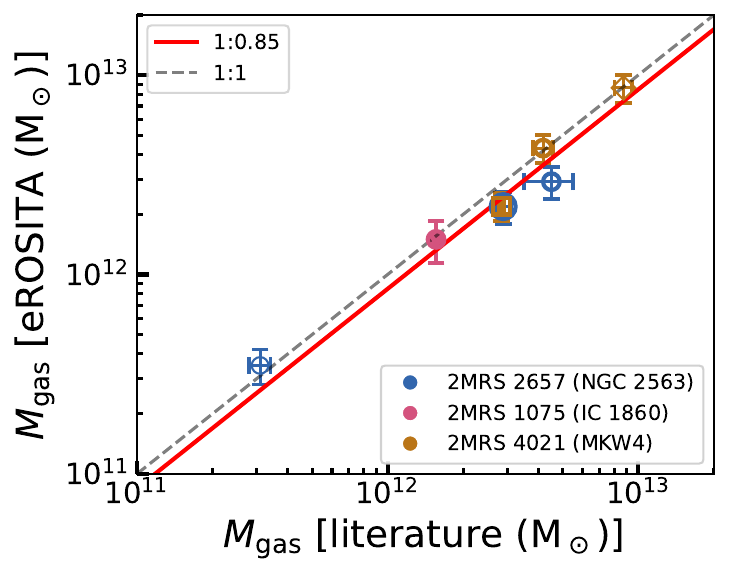}
 \caption{Comparison of gas masses for a matched sample of galaxy groups, comparing our new eROSITA measurements with previous \textit{Chandra} and \textit{XMM-Newton} literature at the same physical radius. eROSITA measures $15\pm4.5\%$ lower $M_{\rm gas}$, in agreement with Fig. 18 in \citet{erass1}.}
\label{comp_lit}
\end{figure}

\section{Sample properties}
In this section we list the optical and X-ray properties of the 2MRS-eRASS1 galaxy group sample studied in this work. In Table \ref{table: data}, we present the coordinates of the X-ray emission centres, redshift, velocity dispersion, core-excised X-ray luminosities, extent of $R_{\rm 500c}$, and number of member galaxies. In Table \ref{table: mass}, we list the total and gas masses of the sample. The best-fit $\beta$ parameter (Sect. \ref{sub:beta_params}) and core radius of the $\beta$-model, together with the mean X-ray background value for each group are listed in Table. \ref{table: beta}.

\begin{table*}
\caption{Optical and X-ray properties of the 2MRS-eRASS1 groups studied in this work.}            
\label{table: data}
\centering                         
\begin{tabular}{c c c c c c c c c c}        
\hline\hline              
Group ID\!\!\!\! &  Field ID &\multicolumn{1}{c}{RA} & \multicolumn{1}{c}{Dec} & \multicolumn{1}{c}{$z$} & \multicolumn{1}{c}{$\sigma$$_{v}$} & $L_{\rm X,500}$ & $L_{\rm X,200}$ & $R_{\rm 500c}$ & $N_{\rm{gal}}$\\
 {\footnotesize 2MRS} & eRASS1 & {\footnotesize (J2000)} & {\footnotesize (J2000)} & & \multicolumn{1}{c}{\footnotesize (km s$^{-1}$)} & (10$^{42}$ erg s$^{-1}$) & (10$^{42}$ erg s$^{-1}$) & (kpc)  & \\
\hline                  
   2938 & 139057 & 139.9323 & 33.7601 & 0.0241 & 430 & $6.21 \pm 0.52$ & $6.85\pm 0.66$ & $559 \pm 65$ & 14\\
   2657 & 126069 & 125.1149 & 21.0893 & 0.0170 & 373 & $2.80 \pm 0.29$ & $3.27\pm 0.33$ & $499 \pm 58$ & 12\\
   7727 & 180069 & 181.0194 & 20.2867 & 0.0248 & 445 & $3.76 \pm 0.43$ & $5.62 \pm 0.63$ & $595 \pm 44$ & 20\\
   4050 & 183066 & 181.9770 & 25.2658 & 0.0230 & 319 & $2.51 \pm 0.37$ & $4.02\pm 0.52$ & $496 \pm 69$ & 10\\
   4808 & 203078 & 202.3497 & 11.7466 & 0.0239 & 347 & $4.36 \pm 0.38$ & $5.91\pm 0.52$ & $471 \pm 45$ & 11\\
   1683 & 068096 & 67.9180 & -5.0859 & 0.0153 & 405 & $2.46 \pm 0.20$ & $2.85\pm 0.28$ & $509 \pm 60$ & 12\\
   3089 & 149096 & 147.4049 & -5.1671 & 0.0220 & 287 & $1.20 \pm 0.25$ & $1.48\pm 0.31$ & $387 \pm 37$ & 9\\
   1075 & 043120 & 42.3849 & -31.2031 & 0.0227 & 292 & $4.99 \pm 0.27$ & $5.76\pm 0.33$ & $373 \pm 60$ & 7\\
   5199 & 215117 & 214.6592 & -27.4093 & 0.0233 & 502 & $6.77\pm 0.56$ & $9.36 \pm 0.81$ & $625 \pm 64$ & 16\\
   343 & 139057 & 16.9808 & -46.8638 & 0.0222 & 364 & $3.43 \pm 0.33$ & $5.76\pm 0.48$ & $438 \pm 62$ & 5\\
   6448 & 287153 & 286.5991 & -62.2351 & 0.0144 & 390 & $0.79 \pm 0.24$ & $1.36\pm 0.30$ & $541 \pm 88$ & 7\\
   391 & 019123 & 18.4809 & -31.7330 & 0.0188 & 284 & $1.15 \pm 0.15$ & $1.63\pm0.21$ & $397 \pm 77$ & 9\\
   4021 & 139057 & 181.1144 & 1.8958 & 0.0211 & 574 & $10.2 \pm 0.55$ & $12.0\pm0.72$ & $684\pm 79$ & 16\\ 
   5318 & 202108 & 200.7273 & -16.9867 & 0.0229 & 233 & $1.32 \pm 0.26$ & $1.96\pm0.36$ & $359 \pm55$ & 10\\
   4957 & 209096 & 207.2674 & -7.2070
   & 0.02401 & 361 & $1.05 \pm 0.24$ & $1.80\pm0.37$ & $444 \pm 64$ & 5\\
   4766 & 201120 & 201.1628 & -30.3003 & 0.01377 & 220 & $0.17 \pm 0.08$ & $0.31\pm0.14$ & $381 \pm 46$ & 11\\
   4757 & 203102 & 201.6672 & -12.3089 & 0.02231 & 329 & $2.12 \pm 0.29$ & $3.26\pm0.41$ & $436 \pm 80$ & 5\\
   4701 & 202105 & 200.2432 & -14.1083 & 0.02192 & 388 & $0.86 \pm 0.27$ & $0.99\pm0.31$ & $430 \pm 76$ & 5\\
   4417 & 194102 & 193.9116 & -12.6915 & 0.01466 & 236 & $0.74 \pm 0.13$ & $0.90\pm0.16$ & $372 \pm 47$ & 11\\
   3807 & 174099 & 174.7969 & -9.3056 & 0.01922 & 261 & $1.17 \pm 0.21$ & $1.24 \pm 0.27$ & $405 \pm 65$ & 6\\
   3657 & 168105 & 168.7264 & -13.6051 & 0.01719 & 364 & $0.75 \pm 0.13$ & -- & $495 \pm 48$ & 5\\
   3224 & 153051 & 153.4901 & 38.8114 & 0.02263 & 404 & $1.10 \pm 0.18$ & $1.40 \pm 0.23$ & $471 \pm 64$ & 11\\
   3148 & 149093 & 149.9206 & -3.1057 & 0.02069 & 329 & $1.02 \pm 0.20$ & $1.25 \pm 0.26$ & $359 \pm 59$ & 7\\
   1925 & 083150 & 79.8134 & -61.1481 & 0.01612 & 252 & $0.75 \pm 0.04$ & -- & $306 \pm 51$ & 5\\
   4271 & 188129 & 188.9642 & -39.9756 & 0.01092 & 310 & $1.52 \pm 0.11$ & -- & $461 \pm 54$ & 9\\

\hline
\end{tabular}
\tablefoot{
RA and Dec are coordinates of the peak large-scale X-ray emission centres. $L_{\rm X}$ is core-excised and computed in the $[0.1-2.4]$ keV band.
}
\end{table*}

\begin{table*}[hbt!]
\caption{Total and gas masses of the 2MRS-eRASS1 groups studied in this work.}             
\label{table: mass}
\centering                          
\begin{tabular}{c c c c c c}       
\hline\hline            
Group ID\!\!\!\! &  Field ID & $M_{\rm 500c}$ & $M_{\rm 200c}$ & $M_{\rm gas,500}$ & $M_{\rm gas,200}$\\
 {\footnotesize 2MRS} & eRASS1 & $10^{13}M_{\odot}$ & $10^{13}M_{\odot}$ & $10^{12}M_{\odot}$ & $10^{12}M_{\odot}$\\
\hline                  
   2938 & 139057 & $5.08\pm1.75$ & $7.52\pm1.88$ & $3.05\pm0.58$ & $6.99\pm0.55$\\
   2657 & 126069 & $3.58\pm1.29$ & $5.27\pm1.38$ & $1.92\pm0.36$ & $3.56\pm 0.41$\\
   7727 & 180069 & $6.14\pm1.36$& $9.01\pm1.39$&$3.17\pm0.44$&$6.41\pm0.56$\\
   4050 & 183066 & $3.54\pm1.53$&$5.25\pm1.64$&$1.88\pm0.49$&$4.23\pm0.49$\\
   4808 & 203078 & $3.03\pm0.88$&$4.41\pm0.92$&$2.19\pm0.38$&$4.55\pm0.48$\\
   1683 & 068096 & $3.83\pm1.38$&$5.63\pm1.48$&$1.75\pm0.34$&$3.43\pm0.40$\\
   3089 & 149096 & $1.68\pm0.62$&$2.48\pm0.64$&$0.80\pm0.15$&$1.40\pm0.26$\\
   1075 & 043120 & $1.51\pm0.76$ & $2.22\pm0.82$&$1.67\pm0.40$&$3.16\pm0.18$\\
   5199 & 215117 & $7.10\pm 2.23$&$10.50\pm2.38$&$3.99\pm0.70$&$7.70\pm1.24$\\
   343 & 139057 & $2.44\pm1.07$&$3.59\pm1.15$&$1.77\pm0.52$&$4.29\pm0.23$\\
   6448 & 287153 & $4.56\pm2.32$&$6.85\pm2.52$&$0.81\pm0.21$&$1.43\pm0.35$\\
   391 & 019123 & $1.81\pm1.03$&$2.69\pm1.14$&$0.81\pm0.2$&$1.51\pm0.24$\\
   4021 & 139057 & $9.29\pm 3.31$&$13.90\pm3.54$&$5.75\pm0.94$&$10.50\pm0.72$\\
   5318 & 202108 & $1.35\pm 0.63$ & $1.97\pm 0.68$ & $0.62\pm 0.18$ & $1.36\pm 0.24$\\
   4957 & 209096 & $2.54\pm 1.13$ & $3.79\pm 1.23$ & $0.79\pm 0.19$ & $1.45\pm 0.35$\\
   4766 & 201120 & $1.60\pm 0.50$ & $2.34\pm 0.62$ & $0.32\pm 0.08$ & $0.54\pm 0.15$\\
   4757 & 203102 & $2.40\pm 1.39$ & $3.59\pm 1.43$ & $1.25\pm 0.47$ & $3.19\pm 0.34$\\
   4701 & 202105 & $2.32\pm 1.29$ & $3.46\pm 1.40$ & $0.60\pm 0.20$ & $1.19\pm 0.34$\\
   4417 & 194102 & $1.49\pm 0.57$ & $2.16\pm 0.61$ & $0.58\pm 0.12$ & $1.07\pm 0.17$\\
   3807 & 174099 & $1.93\pm 0.96$ & $2.86\pm 1.04$ & $0.68\pm 0.17$ & $1.22\pm 0.23$\\
   3657 & 168105 & $3.55\pm 0.98$ & -- & $0.73\pm 0.14$ & --\\
   3224 & 153051 & $3.02\pm 1.26$ & $4.46\pm 1.43$ & $0.88\pm 0.19$ & $1.65\pm 0.29$\\
   3148 & 149093 & $1.34\pm 0.68$ & $1.97\pm 0.74$ & $0.53\pm 0.14$ & $0.91\pm 0.19$\\
   1925 & 083150 & $0.83\pm 0.41$ & -- & $0.36\pm 0.09$ & --\\
   4271 & 188129 & $2.82 \pm 1.0$ & -- & $0.87 \pm 0.22$ & --\\

\hline
\end{tabular}
\end{table*}

\begin{table*}[htbp!]
\caption{SBx profile fitting results of the 2MRS-eRASS1 groups studied in this work.}           
\label{table: beta}
\centering                          
\begin{tabular}{l c c c c}        
\hline\hline                
\!\!\!\!Group ID & Field-ID &  $\beta_{0.15-1\rm R500c}$ & ${R_{\rm C}}$ & ${\rm Bkg}$\ \ \ \ \\

 2MRS & eRASS1 & & ($10^{-3}$ arcmin) & \!\!\!($\log_{10}\rm cts/s/arcmin^{2})$ \\
\hline                  
   2938 & 139057 & $0.41 \pm 0.02$ & $1.52 \pm 0.22$ & -2.765\\ 
   2657 & 126069 & $0.42 \pm 0.03$ & $2.13 \pm 2.95$ & -2.746\\ 
   7727 & 180069 & $0.38 \pm 0.01$ & $0.20\pm 0.42$ & -2.759\\ 
   4050 & 183066 & $0.36\pm 0.03$ & $1.25\pm 0.03$ & -2.707\\ 
   4808 & 203078 & $0.39 \pm 0.02$ & $0.27\pm 0.41$ & -2.361\\ 
   1683 & 068096 & $0.41 \pm 0.03$ & $1.73\pm 1.01$ & -2.580\\ 
   3089 & 149096 & $0.37\pm 0.03$ & $5.14\pm 6.90$ & -2.757\\ 
   1075 & 043120 & $0.46\pm 0.01$ & $945\pm 394$ & -2.814\\ 
   5199 & 215117 & $0.41\pm 0.07$ & $0.28\pm 0.35$ & -2.298\\ 
   343  & 139057 & $0.32\pm 0.01$ & $0.06\pm 1.10$ & -2.714\\ 
   6448 & 287153 & $0.47\pm 0.07$ & $8.02\pm 2.49$ & -2.340\\ 
   391 & 019123 & $0.44\pm 0.04$ & $0.35\pm 0.51$ & -2.746\\  
   4021 & 139057 & $0.47 \pm 0.02$ & $370\pm 30$ & -2.724\\ 
   5318 & 202108 & $0.36 \pm 0.04$ & -- & -2.557\\
   4957 & 209096 & $0.45 \pm 0.06$ & $0.10\pm0.06$ & -2.591\\
   4766 & 201120 & $0.49 \pm 0.11$ & $0.22\pm0.12$ & -2.572\\
   4757 & 203102 & $0.32 \pm 0.01$ & $0.01\pm0.003$ & -2.583\\
   4701 & 202105 & $0.44 \pm 0.08$ & $0.05\pm0.08$ & -2.539\\
   4417 & 194102 & $0.37 \pm 0.02$ & $0.03\pm0.02$ & -2.442\\
   3807 & 174099 & $0.43 \pm 0.02$ & $0.1\pm0.04$ & -2.708\\
   3657 & 168105 & $0.44 \pm 0.04$ & -- & -2.727\\
   3224 & 153051 & $0.38 \pm 0.04$ & $0.01\pm0.01$ & -2.708\\
   3148 & 149093 & $0.34 \pm 0.03$ & $0.08\pm0.04$ & -2.700\\
   1925 & 083150 & $0.43 \pm 0.02$ & $0.16\pm0.004$ & -2.760\\
   4271 & 188129 & $0.23 \pm 0.04$ & -- & -2.898\\
\hline                                
\end{tabular}
\end{table*}

\section{Scaling parameter distributions}
\label{sec: params}
In Figs. \ref{corner1}-\ref{corner_lx_mg200}, we show the 1D and 2D distributions of the posteriors of the scaling relations studied in Sect. \ref{sec: scaling relations}.

\begin{figure}
\centering
\includegraphics[width=0.4\textwidth]{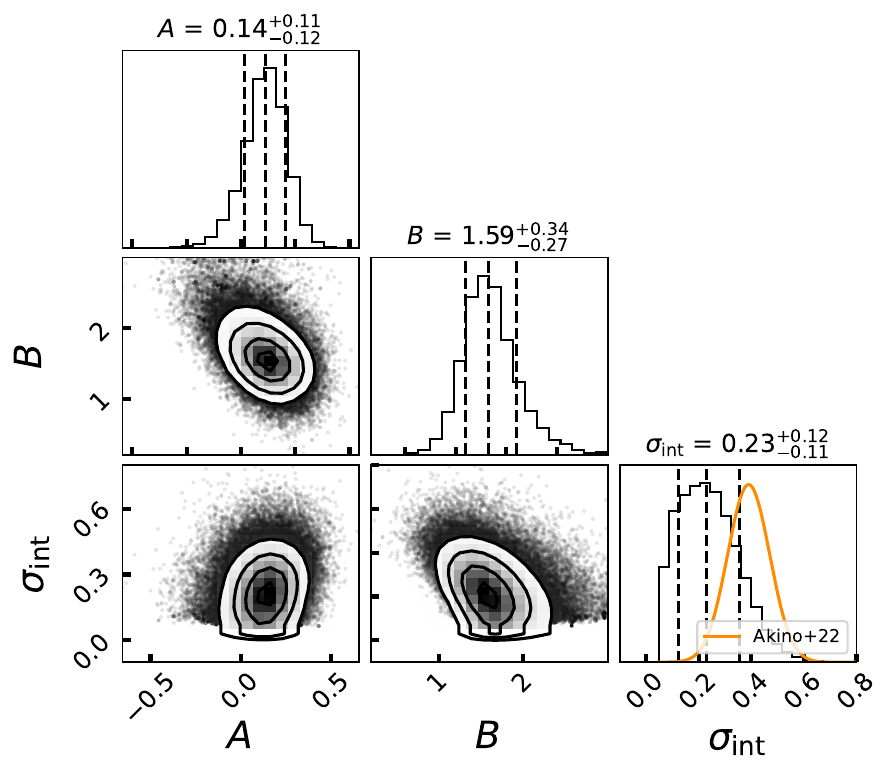}
 \caption{1D and 2D distributions for the posteriors of the $M_{\rm gas,500}-M_{\rm 500c}$ relation parameters according to Eq. \ref{scaling_eqn}. The orange curve shows the intrinsic scatter value from \citet{akino22}.}
\label{corner1}
\end{figure}

\begin{figure}
\centering
\includegraphics[width=0.4\textwidth]{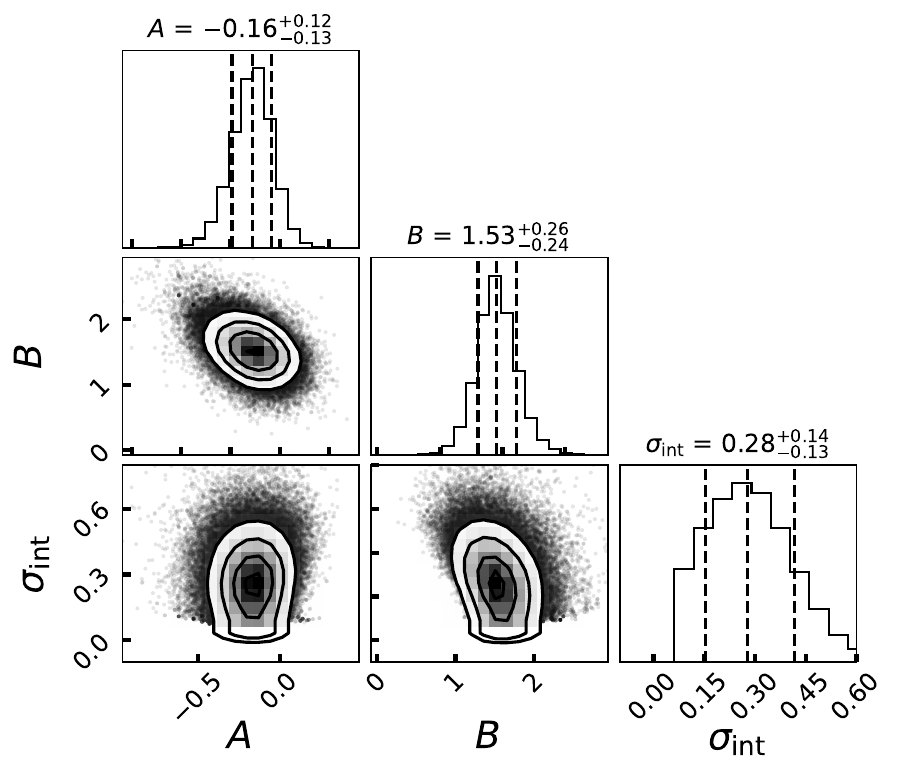}
 \caption{Same as Fig. \ref{corner1} but for $M_{\rm gas,200}-M_{\rm 200c}$.} 
 \label{corner_mg_m200}
\end{figure}

\begin{figure}
\centering
\includegraphics[width=0.4\textwidth]{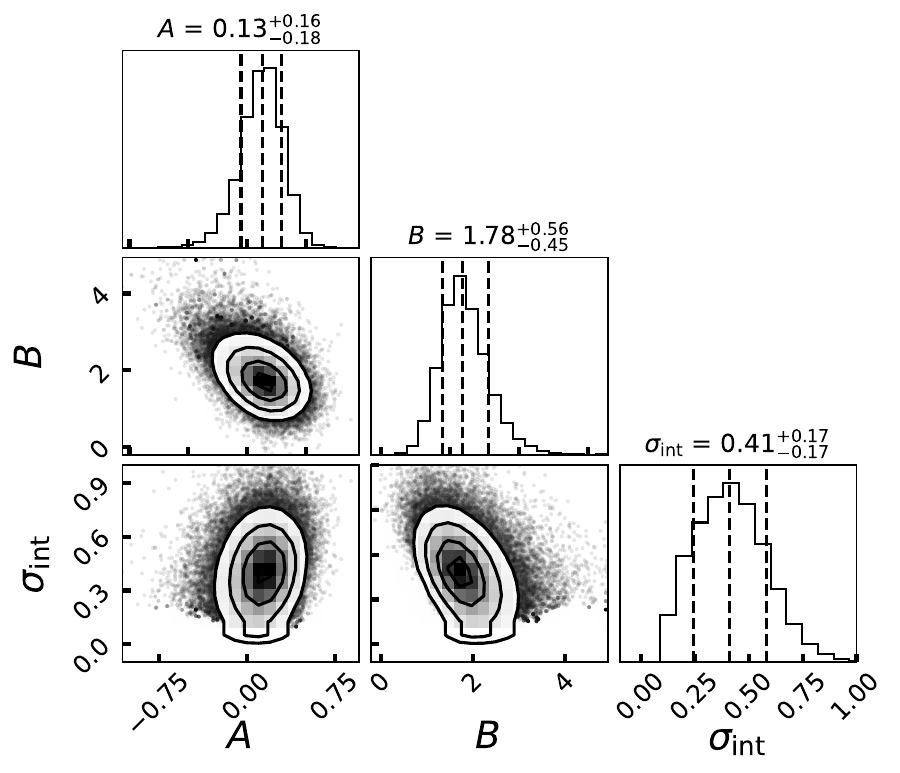}
 \caption{Same as Fig. \ref{corner1} but for $L_{\rm X,500}-M_{\rm 500c}$.} 
 \label{corner_lx_m500}
\end{figure}

\begin{figure}
\centering
\includegraphics[width=0.4\textwidth]{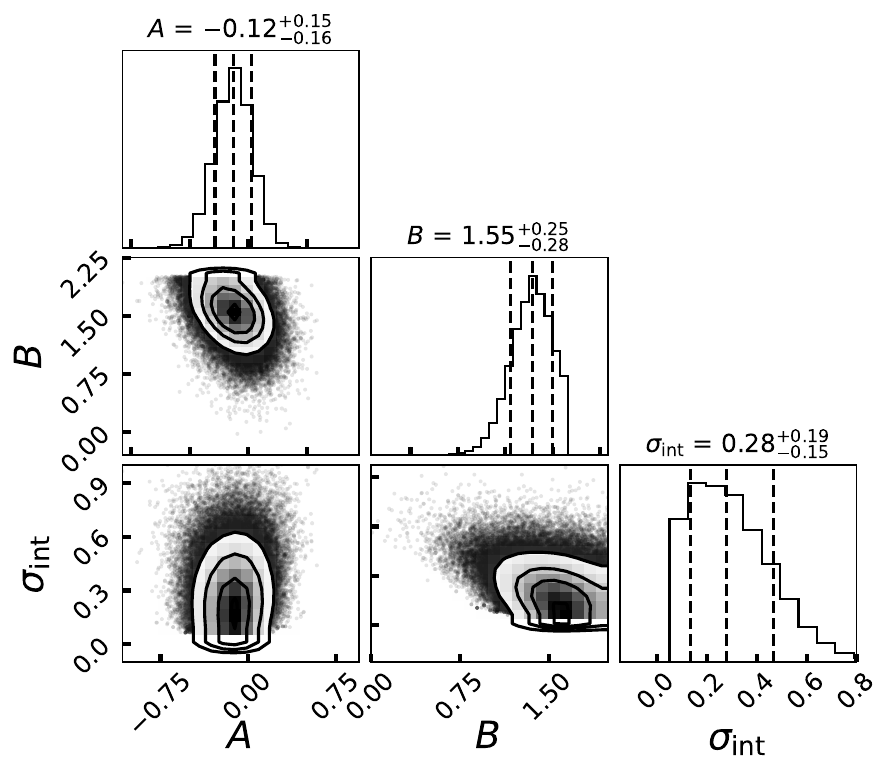}
 \caption{Same as Fig. \ref{corner1} but for $L_{\rm X,200}-M_{\rm 200c}$} 
 \label{corner_lx_m200}
\end{figure}

\begin{figure}
\centering
\includegraphics[width=0.4\textwidth]{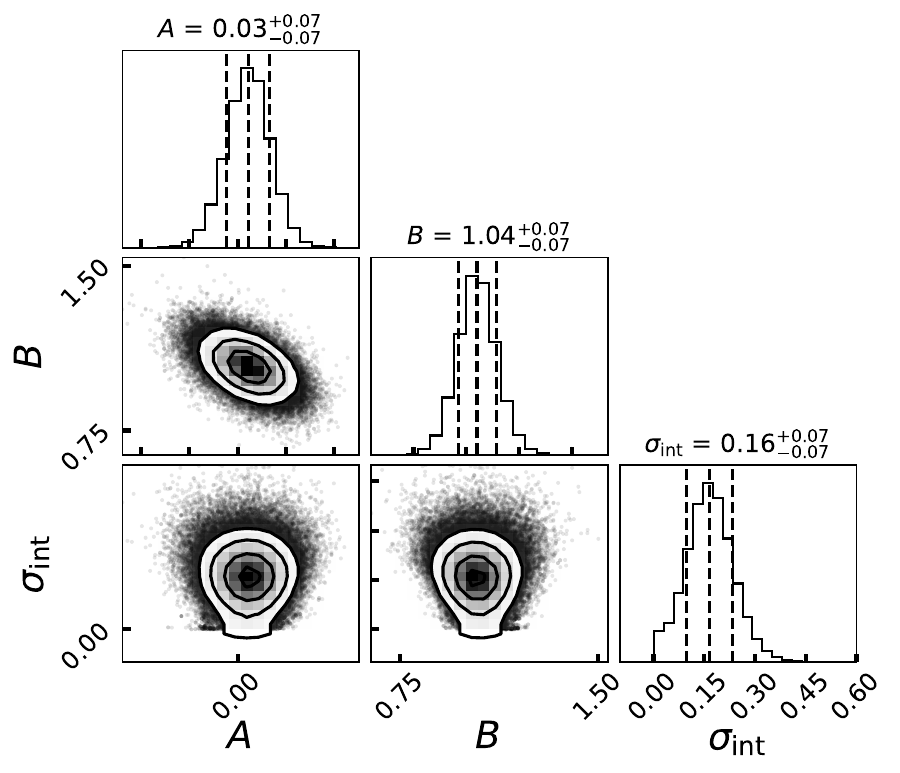}
 \caption{Same as Fig. \ref{corner1} but for $L_{\rm X,500}-M_{\rm gas,500}$} 
 \label{corner_lx_mg500}
\end{figure}

\begin{figure}
\centering
\includegraphics[width=0.4\textwidth]{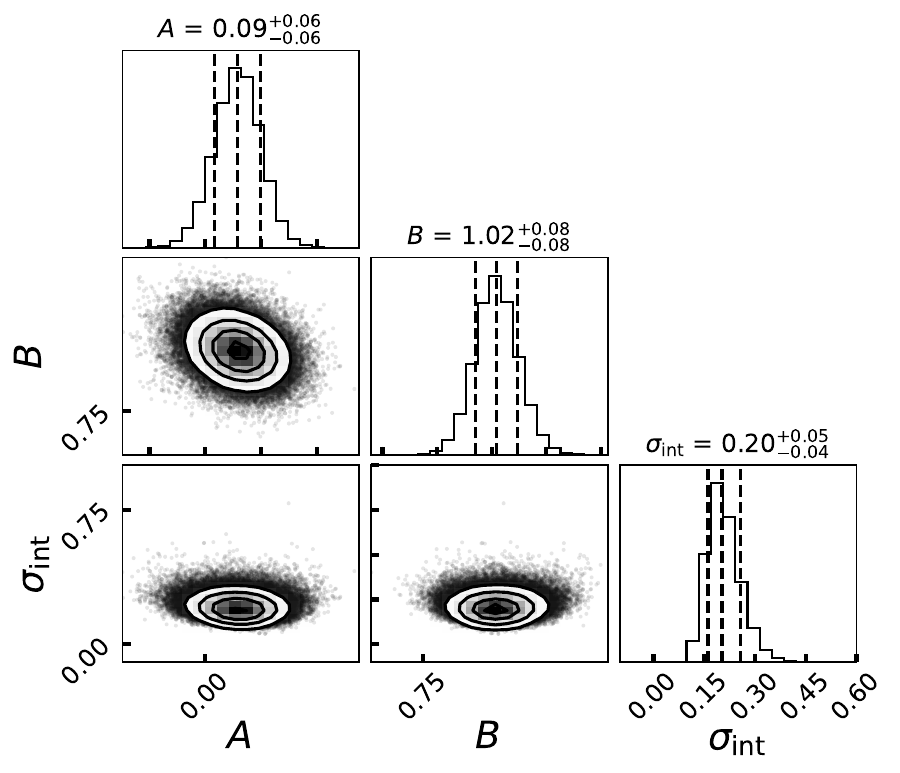}
 \caption{{Same as Fig. \ref{corner1} but for $L_{\rm X,200}-M_{\rm gas,200}$.}} 
 \label{corner_lx_mg200}
\end{figure}

\begin{figure}
\centering
\includegraphics[width=0.4\textwidth]{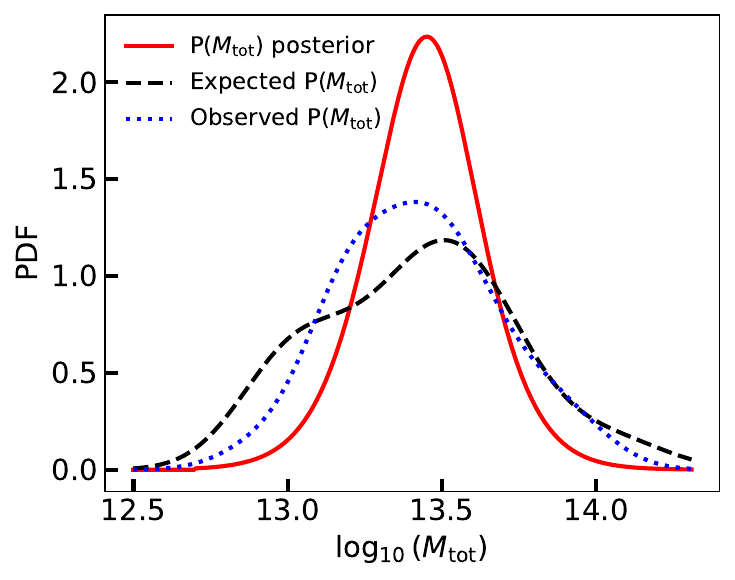}
 \caption{Comparison between the posterior of the distribution of the true masses in our sample (red), the expected true mass distribution inferred from \citet{seppi25}'s modelling (black), and the observed dynamical mass distribution (blue) at $R_{\rm 500c}$.} 
 \label{posterior}
\end{figure}

\section{$L_{\rm X}-M_{\rm gas}$  relation}
\label{sec: lxmg}
X-ray luminosity and gas mass are tightly correlated observables as they both depend on the electron density and trace the hot intragroup medium. We show the $L_{\rm X}-M_{\rm gas}$ relation and provide its best-fit parameter distribution inside $R_{\rm 500c}$ and $R_{\rm 200c}$ in Fig. \ref{lxmgas500} and Table \ref{tab: relations}, respectively.
Our relation at $R_{\rm 500c}$ has a median slope of $B = 1.06 \pm 0.07$ and an intrinsic scatter of $\sigma_{\mathrm{int}} = 0.17\pm0.07$. 

\begin{figure}[htbp!]
    \centering
    \begin{subfigure}[b]{0.8\textwidth}
        \centering
        \hspace{-2.1in}
        \includegraphics[height=0.28\textheight]{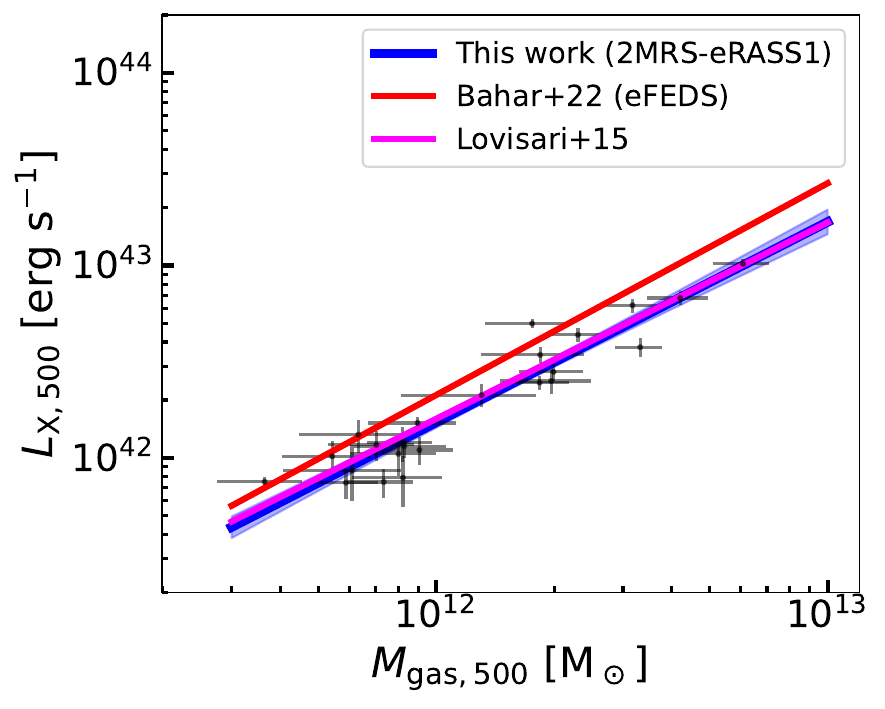}
    \end{subfigure}

    \vspace{0.1cm}

    \begin{subfigure}[b]{0.8\textwidth}
        \centering
        \hspace{-2.1in}
        \includegraphics[height=0.28\textheight]{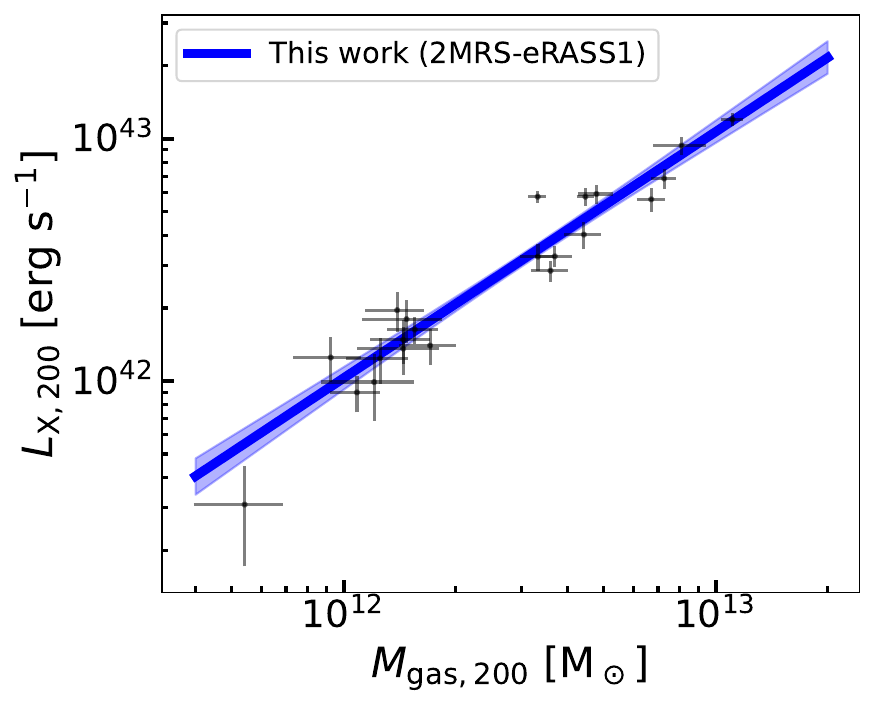}
    \end{subfigure}

    \caption{X-ray luminosity vs gas mass. \textit{Top:} $L_{\rm X}-M_{\rm gas}$ relation inside $R_{\rm 500c}$. The solid blue line shows our best-fit relation for 2MRS-eRASS1. The red and magenta lines show \citet{bahar22}'s eFEDS and \citet{Lovisari15}'s relations, respectively. \textit{Bottom:} $L_{\rm X}-M_{\rm gas}$ relation inside $R_{\rm 200c}$. Other details are the same as in Fig. \ref{mnfwmgas500}.}
    
    \label{lxmgas500}
\end{figure}

We show a comparison to \citet{bahar22} and \citet{Lovisari15}. The former constructed a highly pure sample of 265 groups and clusters in the eFEDS survey, for which they measured eROSITA X-ray properties and constructed the $L_{\rm X}-M_{\rm gas}$ relation, including selection effects. They report a slope of $B = 1.10 \pm 0.025$ consistent with our result and an intrinsic scatter of $\sigma_{\mathrm{int}} = 0.3 \pm 0.02$, substantially higher than our value. We also compared our relation to that of \citet{Lovisari15}, who compiled a complete sample of 20 galaxy groups selected from RASS and observed with \textit{XMM-Newton} with masses of $\sim10^{13} - 10^{14} M_{\odot}$. They report a slope of $B = 1.02 \pm 0.24$, consistent with our relation. We note that in the case of \citet{bahar22}, luminosities are reported in the $[0.5-2.0]$ keV band and were converted to our $[0.1-2.4]$ keV band using a conversion factor of $1.67$ (see Sect. \ref{sec:lxm}). Our results, obtained with the most recent eROSITA software and calibration files, are in agreement with the well-established \textit{XMM-Newton} results of \citet{Lovisari15}, while initial eROSITA results reported by \citet{bahar22} were discordant.

\begin{figure}[H]
\centering
   \includegraphics[width=0.47\textwidth]{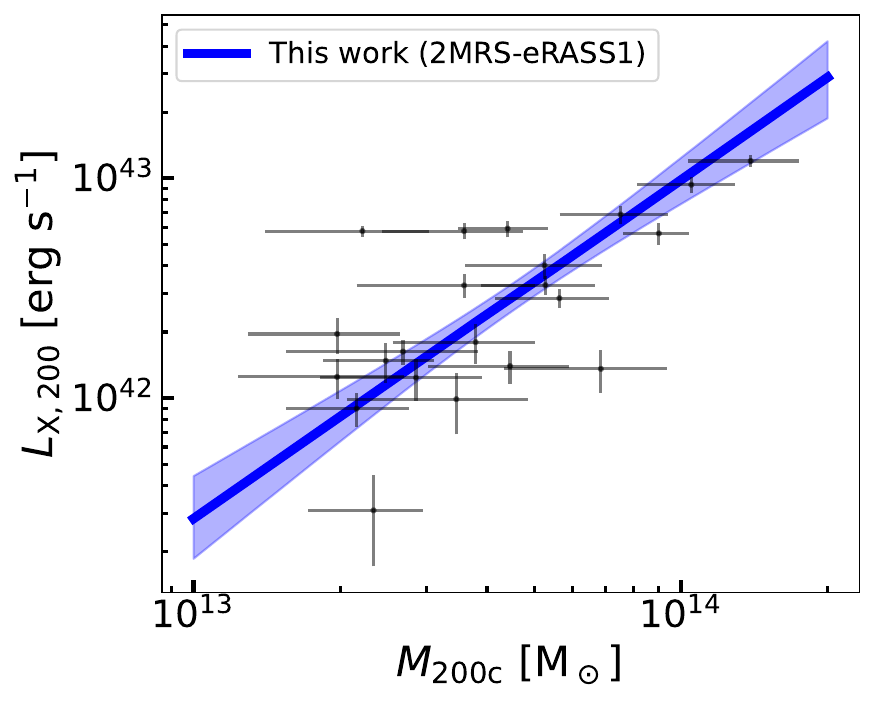}
     \caption{X-ray luminosity vs total mass inside $R_{\rm 200c}$. The solid blue line shows our best-fit relation for 2MRS-eRASS1.} 
     \label{lxm200}
\end{figure}

\section{$L_{\rm X,200c}-M_{\rm tot,200c}$ }
In Fig. \ref{lxm200} we present the X-ray luminosity--total mass scaling relation inside $R_{\rm 200c}$ (see Sect. \ref{sec:lxm} for the relation inside $R_{\rm 500c}$). 

\section{$M_{\rm gas,200c}-M_{\rm tot,200c}$}
In Fig. \ref{mgmnfw200} we present the gas mass--total mass scaling relation inside $R_{\rm 200c}$ (see Sect. \ref{mgas-m} for the relation inside $R_{\rm 500c}$).

\begin{figure}[H]
\centering
   \includegraphics[width=0.47\textwidth]{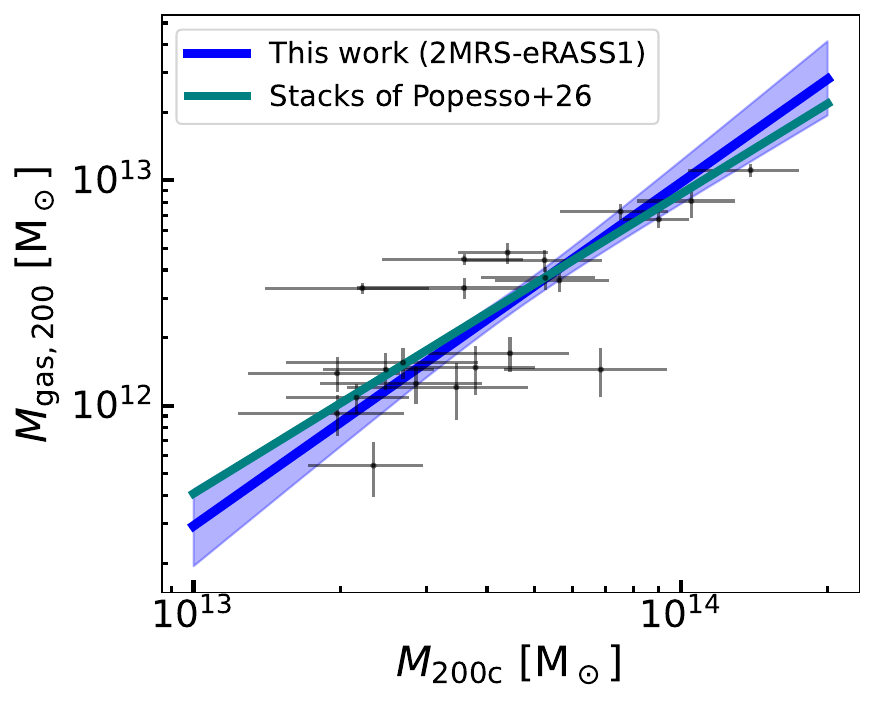}
     \caption{Gas mass vs total mass relation inside $R_{\rm 200c}$. Details are the same as in Fig. \ref{mnfwmgas500}.} 
     \label{mgmnfw200}
\end{figure}

\end{appendix}

\end{document}